\documentclass[12pt]{article}

\usepackage{amsmath}
\usepackage{amssymb}

\usepackage{graphicx}
\usepackage{float}
\usepackage{tikz}
\usetikzlibrary{matrix}
\usepackage{algorithm}
\usepackage{algorithmic}
\usepackage{enumerate}
\usepackage{comment}
\usepackage{subcaption}

\def \E{\mathbb{E}}

\def \cL{{\cal L}}

\def \benu {\begin{enumerate}}
\def \eenu {\end{enumerate}}

\def\beqs{\begin{eqnarray*}}
\def\enqs{\end{eqnarray*}}
\def\beq{\begin{eqnarray}}
\def\enq{\end{eqnarray}}

\newtheorem{defi}{Definition}
\newtheorem{lem}{Lemma}
\newtheorem{theo}{Theorem}
\newtheorem{rem}{Remark}
\newtheorem{pro}{Proposition}
\newtheorem{coro}{Corollary}
\newtheorem{hyp}{Assumption}

\newenvironment{pf}{\ \\ {\bf Proof: }}{\hfill\mbox{$\diamond$}\medskip}

\begin{document}
\vspace{1.5cm}

\author{Erwan Pierre\thanks{EDF R\&D OSIRIS. Email:  \sf erwan.pierre@edf.com}~~~ St\'{e}phane
Villeneuve\thanks{Toulouse School of Economics (CRM-IDEI),
Manufacture des Tabacs, 21, All\'ee de Brienne, 31000 Toulouse,
France.
 Email: \sf
stephane.villeneuve@tse-fr.eu}~~~Xavier Warin\thanks{\small{EDF R\&D \& FiME}, Laboratoire de Finance des
March\'es de l'Energie (www.fime-lab.org)
}
}

\title{Numerical approximation of a cash-constrained firm value with investment opportunities.}
 \vspace{1cm}\renewcommand{\today}

\maketitle

\noindent{\bf Abstract:} We consider a singular control problem with regime switching that arises in problems of optimal investment decisions of cash-constrained firms. The value function is proved to be the unique viscosity solution of the associated Hamilton-Jacobi-Bellman equation. Moreover, we give regularity properties of the value function as well as a description of the shape of the control regions. Based on these theoretical results, a numerical deterministic approximation of the related HJB variational inequality is provided. We finally show that this numerical approximation converges to the value function. This allows us to describe the investment and dividend optimal policies.\\


\section{Introduction}
In a frictionless capital market, Modigliani and Miller theorem demonstrates that firms can fund all valuable investment opportunities. However, if we introduce capital market imperfections,
it is now a standard result that cash-constrained firms have to rely more on internal financial resources: cash holdings and credit line to fund investment opportunities. In recent years, there has been an increasing attention in the use of singular control techniques to model investment problems of a cash-constrained firm.  As references for the theory of singular stochastic control, we may mention the pioneering works of Haussman and Suo \cite{hs:95a} and \cite{hs:95b} and for application to investment/dividend problems Jeanblanc and Shiryaev \cite{js:95},  H\o jgaard and Taksar \cite{ht:99},
Asmussen, H\o jgaard and Taksar \cite{aht:00}, Choulli, Taksar and
Zhou \cite{ctz:03}, Paulsen \cite{pa:03} while more recent studies in corporate finance  include Bolton, Chen and Wang \cite{bcw:11},  D\'ecamps, Mariotti, Rochet and Villeneuve \cite{dmrv:11} and  Hugonnier, Malamud and Morellec \cite{hmm:11}.\\
Singular control is an important class of problems in stochastic control theory. The associated HJB equation, which takes the form of variational inequalities with gradient constraints turns out to be very difficult to solve. In particular, the regularity of the solution are still not well understood. For concrete problems as those arising from dynamic corporate finance, it is thus important to propose a numerical approximation of the value function and to ensure that this numerical approximation converges to the targeted value function.
It turns out from the paper by Barles and Souganidis \cite{bs:91} that when the value function of a singular control problem is the unique viscosity solution of the associated HJB variational inequality, a consistent, stable and monotone numerical scheme converges to the value function.\\
It is now well-established that there exist several approaches to approximate the value function of singular stochastic control problems. First, probabilistic methods based on Markov Chain approximations are essentially explicit finite difference schemes and thus suffer from the stability curse limiting the choice of time step (see \cite{kd:91}). On the other hand, analytical methods based on the tracking of control regions have been developed in \cite{km:04}. They appear to be quite complex because they necessitate a good guess about the shape of control regions especially in the presence of multiple controls.\\
Our paper builds on the theoretical model of cash-constrained firms developed in \cite{pvw:16} with the modification that the investment levels are here discrete. This assumption appears to be reasonable for big industries investing in capacity. The objective is to determine the firm value as well as the investment and dividend optimal policies leading to a singular control problem with regime switching where the regimes correspond to the different levels of production. Our main contributions are
\begin{itemize}
\item We prove that the value function is the unique viscosity solution of the HJB variational inequality. Moreover, we prove the regularity of the value function under a mild assumption about the existence of left and right derivative everywhere. Finally, we prove that it is optimal to pay dividends for high value of cash which allow us to set boundary conditions at right for our numerical scheme.
\item We carry out a rigorous analysis of the direct control method proposed by \cite{hfl:12a}, in the context of HJB variational inequality arising from cash management problem. Having proved a strong comparison theorem, we show that our direct control method is consistent, stable and monotone. In our context, the stability result appears to be a little bit tricky and its proof needs to prove a growth condition on the value function (see Lemma 2).
\item Finally, the numerical approximation of the HJB variational inequality leads to the resolution of a linear system $AU=B$. We present a fixed-point iteration scheme similar to  \cite{hfl:12a} for solving the linear system. To show the convergence of this iterative procedure, we need to prove that the tridiagonal block matrix A is a M-matrix (see Lemma 8) which necessitates an extension of the result proved in \cite{hfl:12a}. 

\end{itemize}

The paper is organized as follows: in section 2, we present the model and derive a standard analytical characterization of the value function in terms of viscosity solutions. Furthermore, we give regularity properties of the value function and a description of the shape of the control regions. Section 3 and 4 are devoted to the presentation of the numerical approximation and contains the convergence result which builds on an extension of the classical techniques developed in \cite{hfl:12a}. Section 5 concludes the paper with numerical illustrations.

\section{The Model}
We consider a firm characterized at each time $t$ by the following balance sheet :\\
$K_t+M_t=L_t+X_t$ where
\begin{itemize}
\item $K_t$ represents the firm's productive assets,
\item $M_t$ represents the amount of cash reserves or liquid assets,
\item $L_t$ represents the volume of outstanding debt,
\item $X_t$ represents the book value of equity.
\end{itemize}

We suppose that the firm is able to choose the level of its productive assets, by investment or disinvestment, in a range of $N$ strictly positive levels :
$
K_t \in \left\{k_i\right\}_{i \in [1,N]}.
$  
This assumption appears to be reasonable for big industries  for which increasing the productive assets involves large sunk cost as the construction of new plant or the purchase of new equipments. Without loss of generality, we suppose that $\left\{k_i\right\}_{i \in [1,N]}$ satisfy 
\[
\forall i \in [1,N], k_i = k_1 + (i-1)h.
\]
The productive assets continuously generate cash-flows $(R_t)_{t\ge 0}$ over time. We assume 
\[
dR_t = \beta(K_t)(\mu dt + \sigma dB_t),
\] 
where $\mu$ and $\sigma$ are positive constants and $(B_t)_{t\ge 0}$ is a standard Brownian motion on a complete probability space $(\Sigma, \mathcal{F}, \mathbb{P})$ equipped with a filtration $(\mathcal{F}_t)_{t\geq 0}$. In this model, we assume a decreasing-return-to-scale technology by introducing the increasing bounded concave function $\beta$. We denote $\bar{\beta}$ the maximum taken by the gain function on $\{k_i\}_{i\in[1,N]}$ i.e.
\[
\bar{\beta} = \beta(k_N).
\]
In order to finance its working capital requirement, we consider that the firm has access to a secured credit line. The collateral of the credit line is given by the market value of the firm assets. If we introduce $\gamma > 0$ the cost to disinvest the productive assets and $M$ the level of cash, the credit line's depth is assumed to be
\begin{equation}\label{max_debt}
L_{\max} = (1 - \gamma) K + M.
\end{equation}
When this credit line limit is reached, the company is no longer able to meet its financial commitments and is therefore forced to go bankrupt. At this point, the manager liquidates the firm assets in order to refund the creditors with priority for debt holders over shareholders. In order to make the credit line attractive for bank's shareholders, we make the following assumption :

\begin{hyp}\label{hyp_det}
$\alpha$ is a strictly continuously differentiable function. Furthermore, it is assumed that the collateralized debt is profitable for the bank meaning that
\[
\forall x \geq 0, \alpha'(x) > r \text{ and } \alpha(0) = 0.
\]
\end{hyp}


In \cite{pvw:16}, it has been proved in a similar framework that it is optimal to use the credit line if and only if the cash reserves are depleted meaning that 
\begin{equation}\label{optimal_debt}
L_t = (K_t - X_t)^+.
\end{equation}
We therefore have the following dynamics for the book value of equity and the productive assets (\cite{pvw:16}):
\[
\left\{
\begin{split}
dX_t &= \beta(K_t)(\mu dt + \sigma dB_t) - \alpha((K_t - X_t)^+)dt - \gamma |dI_t| - dZ_t\\
dK_t &= dI_t^+ - dI_t^-,
\end{split}
\right.
\]
where $Z = (Z_t)_{t \geq 0}$ is an increasing right-continuous $(\mathcal{F}_t)_t$ adapted process representing the cumulative dividend payments up to time $t$ and $I^+ = (I^+_t)_{t\geq 0}$ (respectively $I^-$) is the cumulative investment process (respectively disinvestment). Here we suppose that the cost to investment is the same as the cost of disinvestment $\gamma$.\\
The manager acts in the interest of the shareholders and maximizes the expected discounted value of all future dividend payout. Shareholders are assumed to be risk-neutral and future cash-flows are discounted at the risk-free rate $r$. Thus, the objective is to maximize over the admissible control $\pi = (I^+,I^-,Z)$ the functional
\[
V(x,k_i;\pi) = \mathbb{E}_{x,k_i}\left(\int_0^\tau e^{-rt}dZ^\pi_t\right),
\]
where $x$ and $k_i$ are the initial values of equity capital and productive capital. $\tau$ is the time of bankruptcy and according to (\ref{max_debt}) and (\ref{optimal_debt}), we have
\[
\tau = \inf_{t\geq 0}\{X^\pi_t \leq \gamma K^\pi_t\}.
\]
We denote by $\Pi$ the set of admissible control variables and define the shareholders value functions by
\[
\forall i \in [1,N], v_i(x) = v(x,k_i) = \sup_{\pi \in \Pi} V(x,k_i;\pi),
\]
which are defined on the domains
\[
\forall i \in [1,N], \Omega_i= [\gamma k_i, +\infty[.
\]

\subsection{Viscosity solutions}

The aim of this section is to determine the HJB variational inequality (HJB-VI) satisfied by the shareholders value functions $(v_i)_{i\in[1,N]}$. This analytical characterization will allow us to solve numerically the problem of optimal investment for a cash-constrained firm.

\begin{pro}\label{cont}
The shareholders value functions $v_i$ are jointly continuous for every $i \in[1,N]$.
\end{pro}
\begin{pf}
Take $i\in[1,N]$, $x > \gamma k_i$ and $(x_n)_{n\in \mathbb{N}}$ a sequence of $\Omega_i$ that converges to $x$. We consider two admissible strategies :
\begin{itemize}
\item Strategy $\pi_n^1$ : from $(x,k_i)$ wait until the exit time of the interval $(\gamma k_i,x_n)$. We denote $(X^{\pi_n^1}_t,K^{\pi_n^1}_t)_{t\geq 0}$ the process controlled by $\pi_n^1$.
\item Strategy $\pi_n^2$ : from $(x_n,k_i)$ wait until the exit time of the interval $(\gamma k_i,x)$ We denote $(X^{\pi_n^2}_t,K^{\pi_n^2}_t)_{t\geq 0}$ the process controlled by $\pi_n^2$.
\end{itemize}
We define
\[
\theta^1_n = \inf\{t\geq 0, (X_t^{\pi_n^1}, K_t^{\pi_n^1}) = (x_n,k_i)\},\,
\theta^2_n = \inf\{t\geq 0, (X_t^{\pi_n^2}, K_t^{\pi_n^2}) = (x,k_i)\},
\]
and
\[
T^1_n = \inf\{t\geq 0, X_t^{\pi_n^1} = \gamma K_t^{\pi_n^1})\},\,
T^2_n = \inf\{t\geq 0, X_t^{\pi_n^2} = \gamma K_t^{\pi_n^2})\}.
\]
Dynamic programming principle and $v(X_{T^1_n},K_{T^1_n}) = 0$ yield
\[
\begin{split}
v_i(x) &\geq \mathbb{E}\left[\int_0^{\theta^1_n \wedge T^1_n}e^{-r t}dZ^{\pi^1_n}_t + e^{-r(\theta^1_n \wedge T^1_n)} 1_{\{\theta^1_n < T^1_n\}}v(X_{\theta^1_n},K_{\theta^1_n})\Big) \right]\\
&\geq \mathbb{E}\left[e^{-r\theta^1_n} 1_{\{\theta^1_n < T^1_n\}}v_i(x_n)\right]\\
&\geq \left(\mathbb{E}\big(e^{-r\theta^1_n}\big) - \mathbb{E}\big(e^{-r\theta^1_n}1_{\{\theta^1_n \geq T^1_n\}}\big)\right)v_i(x_n)\\
&\geq \left(\mathbb{E}\big(e^{-r\theta^1_n}\big) - \mathbb{P}\big(\theta^1_n \geq T^1_n\big)\right)v_i(x_n).\\
\end{split}
\]
On the other hand, using $v(X_{T^2_n},K_{T_n^2})=0$ we have 
\[
\begin{split}
v_i(x_n) &\geq \mathbb{E}\left[\int_0^{\theta^2_n \wedge T^2_n}e^{-r t}dZ^{\pi^2_n}_t + e^{-r(\theta^2_n \wedge T^2_n)} 1_{\{\theta^2_n < T^2_n\}}v(X_{\theta^2_n},K_{\theta_n^2})\Big) \right]\\
&\geq \mathbb{E}\left[e^{-r\theta^2_n} 1_{\{\theta^2_n < T^2_n\}}v_i(x)\right]\\
&\geq \left(\mathbb{E}\big(e^{-r\theta^2_n}\big) - \mathbb{E}\big(e^{-r\theta^2_n}1_{\{\theta^2_n \geq T^2_n\}}\big)\right)v_i(x)\\
&\geq \left(\mathbb{E}\big(e^{-r\theta^2_n}\big) - \mathbb{P}\big(\theta^2_n \geq T^2_n\big)\right)v_i(x).\\
\end{split}
\]
In \cite{pvw:16}, the authors proved that
\[
\begin{split}
&\lim_{n \rightarrow \infty}\mathbb{P}(\theta_n^1 \geq T_n^1) = 0,\\
&\lim_{n \rightarrow \infty}\mathbb{P}(\theta_n^2 \geq T_n^2) = 0.
\end{split}
\]
and
\[
\lim_{n \rightarrow +\infty} \mathbb{E}(e^{-r\theta_n^1}) = \lim_{n \rightarrow +\infty} \mathbb{E}(e^{-r\theta_n^2}) = 1.
\]
Then
\[
v_i(x)\ge \limsup_n v_i(x_n) \ge \liminf_n v_i(x_n) \ge v_i(x),
\]
which proves the continuity of $v_i$ on the open interval $(\gamma k_i,+\infty)$. It remains to show the continuity at the boundary point $\gamma k_i$. Using integration by part, we have for $x>\gamma k_i$ and a fixed strategy $\pi$,
$$
\E_x\left[e^{-r\tau}X_{\tau}^\pi\right] -x= \E_x\left[\int_0^\tau e^{-rs}(dX^\pi_s-rX^\pi_s\,ds)\right].
$$
Because $X^\pi$ is always nonnegative, we obtain 
\begin{align*}
\E_x\left[\int_0^\tau e^{-rs}dZ_s^\pi \right] &\le x-\E_x\left[e^{-r\tau}X_{\tau}^\pi\right]+ \\
& \E_x\left[\int_0^\tau e^{-rs}(\beta(K_t)(\mu dt + \sigma dB_t) - \alpha((K_t - X_t)^+)dt - \gamma |dI_t|\right]\\
&\le \frac{\bar{\beta}\mu}{r}\left(1-\E_x\left[e^{-r\tau}\right]\right)+x-\E_x\left[e^{-r\tau}X_{\tau}^\pi\right].
\end{align*}
As disinvestment moves the pair equity-productive assets parallel of the liquidation boundary and as we cannot disinvest at the level of productive assets $k_1$, the liquidation boundary is crossed continuously. Therefore, we can assume without loss of generality that we cross the level $\gamma k_i$ continuously. Letting $x$ tend to $\gamma k_i$, the right-hand side converges to zero because $\gamma k_i$ is a regular point for $X^\pi$, proving the continuity of the value function at $\gamma k_i$.
\end{pf}

Let $\mathcal{L}_i$ be the next differential operator:
\begin{equation}
\mathcal{L}_i\phi = (\beta(k_i)\mu - \alpha((k_i-x)^+))\phi'(x) + \frac{\beta(k_i)^2\sigma^2}{2}\phi''(x) - r \phi(x).
\end{equation}
The next lemma establishes a well-known comparison principle which we shall use to prove a linear growth condition for the shareholders value function. The proof is omitted.
\begin{lem}\label{comparison}
  Suppose $(\varphi_i)_{i \in [1,N]}$ are $N$ smooth functions on $(\gamma k_i, +\infty)$ such that $\varphi_i(\gamma k_i) \geq 0$ and
  \begin{equation}\label{viscc}
  \forall i \in [1,N], \forall x \geq \gamma k_i, \min\left[-\mathcal{L}_i \varphi_i(x), \varphi_i'(x) - 1, \varphi_i(x) - \max_{j \neq i} \varphi_j(x - \gamma |k_i - k_j|) \right] \geq 0,
  \end{equation}
  then we have for all $i\in [1,N]$, $v_i \leq \varphi_i$.
  \end{lem}

As a corollary, we prove a linear growth condition for the shareholders value functions $v_i$.
\begin{lem}\label{linear_growth}
For all $i \in [1,N]$ and for all $x \in \Omega_i$, we have
\[
v_i(x) \leq x - \gamma k_i + \frac{\mu \bar{\beta}}{r}.
\]
\end{lem}
\begin{pf}
For all $i \in [1,N]$, we define
\[
\varphi_i(x) = x - \gamma k_i + \frac{\mu \bar{\beta}}{r}.
\]
We prove easily that $(\varphi_i)_{i \in [1,N]}$ are viscosity supersolutions of (\ref{viscc}). Indeed,
\[
\forall i \in [1,N], \varphi_i'(x) \geq 1,
\]
and
\[
\varphi_i(x) - \varphi_j(x - \gamma |k_i - k_j|) = \gamma |k_i - k_j| - \gamma (k_i - k_j) \geq 0,
\]
and
\[
-\mathcal{L}_i \varphi_i(x) = - (\beta(k_i) \mu - \alpha((k_i - x)^+)) + r ( x - \gamma k_i) + \mu \bar{\beta} \geq 0,
\]
and we have $\varphi_i(\gamma k_i) = \frac{\mu \bar{\beta}}{r} > v_i(\gamma k_i)$ using that $v_i(\gamma k_i)=0$. Lemma \ref{comparison} proves the result.
\end{pf}

We are now in a position to state the main result of this section.

\begin{pro}\label{sol_visc}
The shareholders value functions $(v_i)_{i\in [1,N]}$ are the unique continuous viscosity solutions to the HJB variational inequality :
\begin{equation}\label{visc_eq}
\begin{split}
&\forall i \in [1,N], \forall x \geq \gamma k_i,\\
&\min\left\{-\mathcal{L}_iv_i(x), v_i'(x) - 1, v_i(x) - \max_{j\neq i} v_j(x-\gamma|k_i - k_j|) \right\} = 0\\
\end{split}
\end{equation}
with boundary conditions
\begin{equation} \label{defaultboundary}
\forall i \in [1,N], v_i(\gamma k_i) = 0.
\end{equation}
\end{pro}

\begin{pf}
The proof  is very similar to the one proved in \cite{pvw:16} and thus omitted.
\end{pf}

\begin{rem}\label{rem_10}
It is sufficient to impose the boundary condition \eqref{defaultboundary} to have the uniqueness of the viscosity solution.
\end{rem}

\begin{rem}\label{rem_1}
Because $v_i(x) \geq v_j(x-\gamma |k_i-k_j|)$ for all pairs $(i,j)$, the HJB-VI is equivalent to
\[
\begin{split}
&\forall i \in [1,N], \forall x \geq \gamma k_i,\\
&\min\Big\{-\mathcal{L}_iv_i(x), v_i'(x) - 1, v_i(x) - \max\Big(v_{i-1}(x-\gamma h), v_{i+1}(x-\gamma h)\Big)\Big\} = 0.\\
\end{split}
\]
\end{rem}

We end this section by giving heuristic arguments to understand HJB-VI \eqref{visc_eq}. Assume the firm has $k_i$ level of productive assets. In the region where it is optimal  to retain cash, the value function satisfies $\cL v_i=0$ explaining the first term. For the second term, assume the firm pays a dividend $\varepsilon$ at time 0. Because this strategy is a priori suboptimal, we have 
$v_i(x)=v_i(x-\varepsilon)+\varepsilon$ yielding $v_i^{'}\ge 1$. For the last term, assume the firm invest (resp. disinvest) to switch from $k_i$ to $k_j$ with $k_i <k_j$ (resp. $k_i>k_j$). Therefore, for every $j$, we have $v_i(x)\ge v_j(x-\gamma \vert k_i-k_j\vert)$.

\subsection{Regularity}

We set for all $i \in [1,N]$,
\[
\begin{split}
  \mathcal{S}_{ij} &= \{ x \in \Omega_i, v_i(x) = v_j(x - \gamma |k_i - k_j|) \},\\
  \mathcal{S}^+_i &= \cup_{j>i}\mathcal{S}_{ij},\\
  \mathcal{S}^-_i &= \cup_{j<i}\mathcal{S}_{ij},\\
  \mathcal{S}_i &= \mathcal{S}_i^+ \cup \mathcal{S}_i^-,\\
\end{split}
\]
which define the investment region ($\mathcal{S}^+_i$) and the disinvestment one ($\mathcal{S}^-_i$).

Before proving the main result of this section, we need to establish some preliminary results on the value function.

\begin{lem}\label{sup_un}
  $\forall i\in [1,N]$, $v_i(x+h) \geq v_i(x) + h$.
\end{lem}
\begin{pf}
  It is obvious if we consider the sub-optimal strategy from initial state $(k_i, x + h)$ which consists on distributing $h$ dollars as dividends at time $t=0$ and following the optimal strategy hereafter. By the dynamic programming principle we have : 
  \[
  v_i(x+h) \geq v_i(x) + h.
  \]
\end{pf}
\begin{lem}\label{lem_invdesinv}
  $\forall i \in [1,N]$,
  \begin{enumerate}
  \item $\forall x \in \mathcal{S}^+_i, x - \gamma h \notin \mathcal{S}^-_{i+1}$.
  \item  $\forall x \in \mathcal{S}^-_i, x - \gamma h \notin \mathcal{S}^+_{i-1}$.
  \end{enumerate}
\end{lem}
\begin{pf}
  We prove only the first assertion  since the demonstration of the second assertion is similar. Suppose on the contrary that there exists $i\in [1,N]$ and $x\in S_i^+$ such that $x - \gamma h \in S^-_{i+1}$. Therefore,
  \[
  v_i(x) = v_{i+1}(x - \gamma h) = v_{i}(x - 2\gamma h)
  \]
  which is in contradiction with lemma \ref{sup_un}.
\end{pf}\\
We proved in the previous section that the value function is continuous. It has been proved in \cite{pham:springer} that the convexity of the value function is a sufficient condition for the regularity of its derivative. But, as proved in \cite{pvw:16}, the value function might be convex-concave when the credit line interest rate is high. Nonetheless, we will give below a regularity result under the following assumption about the existence of left and right derivatives.

\begin{hyp}\label{rlderiv}
The value function admits left ($D^-$) and right ($D^+$) derivatives on its definition domain.
\end{hyp}
\begin{pro}
  Under the assumption (\ref{rlderiv}), the value functions $v_i$ are $C^1$ for all $i\in [1,N]$.
\end{pro}
\begin{pf}
  Let be $i\in[1,N]$, $x_0 \in \Omega_i$ and suppose $D^+v_i(x_0) > D^-v_i(x_0)$. Then take some $q \in (D^-v_i(x_0), D^+v_i(x_0))$ and consider the function
  \[
  \varphi_i(x) = v_i(x_0) + q(x-x_0) + \frac{1}{2\epsilon}(x-x_0)^2
  \]
  with $\epsilon > 0$. Then $x_0$ is a local minimum of $v_i-\varphi_i$, with $\varphi'(x_0) = q$ and $\varphi''(x_0) = \frac{1}{\epsilon}$. Therefore, we get a contradiction by writing the supersolution inequality :
  \[
  0 \leq -(\beta(k_i) \mu - \alpha((k_i - x)^+))q + r(v_i(x_0)) - \frac{\sigma^2\beta(k_i)^2}{2 \epsilon} 
  \]
  and choosing $\epsilon$ small enough. So we have the inequality
  \[
  D^+v_i(x_0) \leq D^-v_i(x_0).
  \]\\
  Suppose now that there exists some $x_0 \notin \mathcal{S}_i$ such that $D^-v_i(x_0) > D^+v_i(x_0)$. We then fix some $q \in (D^+v_i(x_0), D^-v_i(x_0))$ and consider the function
  \[
  \varphi(x) = v_i(x_0) + q(x-x_0) - \frac{1}{\epsilon}(x-x_0)^2
  \]
  with $\epsilon > 0$. Then $x_0$ is a local maximum of $v_i- \varphi$ with $\varphi'(x_0) = q > D^+v_i \geq 1$. Since $x_0 \notin \mathcal{S}_i$, the subsolution inequality property implies
  \[
  0 \geq r v_i(x_0) - (\mu \beta(k_i) - \alpha((k_i - x)^+))q + \frac{\sigma^2\beta(k_i)^2}{2 \epsilon}
  \]
  which leads to a contradiction by choosing $\epsilon$ sufficiently small. Therefore, we have that $v_i$ is $C^1$ on the open set $\Omega_i\backslash \mathcal{S}_i$.\\
  Let's prove now that $v_i$ is still $C^1$ on $\mathcal{S}_i$. Fix $x_0 \in \mathcal{S}^+_i$ (the proof for $x_0 \in \mathcal{S}^-_i$ is similar) and take $j = \min\{l > i, x_0 - \gamma |k_i - k_l| \notin \mathcal{S}^+_l\}$. Then $x_0$ is a minimum of $v_i - v_j(. - \gamma (k_j - k_i))$, and so
  \[
  D^-v_i(x_0) - D^-v_j(x_0 - \gamma (k_j - k_i)) \leq D^+v_i(x_0) - D^+v_j(x_0 - \gamma (k_j - k_i)).
  \]
  But, from the definition of $j$, $x_0 - \gamma(k_j - k_i) \notin S^+_j$ and from Lemma \ref{lem_invdesinv}, $x_0 - \gamma(k_j - k_i) \notin S^-_j$ so $x_0 - \gamma(k_j - k_i)$ belongs to the open set $\Omega_j \backslash S_j$ and so $D^+v_j(x_0-\gamma (k_j - k_i)) =  D^-v_j(x_0-\gamma (k_j - k_i))$ and thus
  \[
  D^-v_i(x_0) \leq D^+v_i(x_0)
  \]
  which proves the result since the reverse inequality has been proved previously.
\end{pf}

Since we prove, under the Assumption \ref{rlderiv}, that the value function is $C^1$, we pose from now on : 
\[
\begin{split}
\mathcal{D}_i &= \{x \in \Omega_i, v'_i(x) = 1\},\\
\mathcal{C}_i &= \Omega_i\backslash (\mathcal{D}_i \cup \mathcal{S}_i).\\ 
\end{split}
\]

\begin{pro}\label{derivC2}
  For all $i\in [1,N], v_i$ is $C^2$ on $\mathcal{C}_i$.
\end{pro}

\begin{pf}
In this open set, we have that $v_i$ is a viscosity solution to
  \begin{equation}\label{deriv22}
  -\mathcal{L}_i v_i = 0, x \in \mathcal{C}_i.
  \end{equation}
  Now, for any arbitrary bounded interval $(x_1, x_2) \in \mathcal{C}_i$ consider the Dirichlet boundary linear problem :
  \begin{equation}\label{deriv2}
  \begin{split}
  &-\mathcal{L}_i w = 0\\
  &w(x_1) = v_i(x_1), \qquad w(x_2) = v_i(x_2).
  \end{split}
  \end{equation}
  Classical results (see for instance \cite{fr:75}) provide the existence and uniqueness of a smooth $C^2$ function $w$ solution on $(x_1,x_2)$ to (\ref{deriv2}). In particular, this smooth function $w$ is a viscosity solution to (\ref{deriv22}). From standard uniqueness results, we get $v_i = w$ on $(x_1,x_2) \in \mathcal{C}_i$ which proves that $v_i$ is $C^2$ on $\mathcal{C}_i$ from the arbitrariness of $(x_1,x_2)$.
\end{pf}

\subsection{Properties of the dividend region}\label{div_reg}

At this point, we only have the boundary condition :
\[
\forall i \in [1,N], v_i(\gamma k_i) = 0.
\]
However, to solve numerically the problem, we need another boundary condition on the right side.
The next lemma gives us a property of the dividend region that will make the numerical scheme well-posed.\\ 

\begin{lem}\label{lem_bi_finit}
  For all $i \in [1,N]$, we have
  \[
  b_i = \sup\{x \in \Omega_i, v'_i(x) > 1\} < +\infty.
  \]
\end{lem}
\begin{pf}
  We note $\mathbb{I} = \{ i \in [1,N], b_i < +\infty\}$ and we suppose that $\mathbb{I}^c = [1,N] \backslash \mathbb{I} \neq \emptyset$. For all $i\in [1,N]$, the function $x \rightarrow v_i(x) - x$ is an increasing bounded continuous function (see lemma \ref{linear_growth}) and therefore admits a limit $a_i = \lim_{x \rightarrow +\infty}(v_i(x) - x)$. We have for all $(i,j) \in [1,N] \times [1,N]$ :
  \[
  a_j - (a_i - \gamma|k_i - k_j|) = \lim_{x \rightarrow +\infty}(v_j(x) - v_i(x - \gamma |k_i - k_j|)) \geq 0.
  \]
  Take $j_0 \in \mathbb{I}^c$ such that $a_{j_0} = \max \{a_j, j \in \mathbb{I}^c\}$. In particular, we have for all $j \in \mathbb{I}^c \backslash \{j_0\}, a_{j_0} > (a_j - \gamma |k_{j_0}-k_j|)$. We prove easily that there exists $\bar{x} \in \mathbb{R}^+$ such that
  \[
  \left\{
  \begin{split}
    &\bar{x} > k_{j_0},\\
    &r v_{j_0}(\bar{x}) > \mu \beta(k_{j_0}),\\
    &v_{j_0}(\bar{x}) > \bar{x} + \max_{j \in \mathbb{I}^c \backslash \{j_0\}}(a_j - \gamma |k_{j_0}-k_j|),\\
    &\bar{x} > b_i + \gamma |k_i - k_{j_0}|, \forall i \in \mathbb{I}.
  \end{split}
  \right.
  \]
  We then define the function $w$ such that
  \[
  \begin{split}
    \forall x \leq \bar{x}, w(x) = v_{j_0}(x),\\
    \forall x > \bar{x}, w(x) = v_{j_0}(\bar{x}) + x - \bar{x}.
  \end{split}
  \]
  Then by definition, for $x\in [\gamma k_{j_0}, \bar{x}]$, $w$ is a viscosity solution of
  \[
  \min\left\{ - \mathcal{L}_{j_0}w, w'(x) - 1, w(x) - \max_{j \neq j_0}v_j(x - \gamma|k_{j_0} - k_j|)\right\} = 0.
  \]
  We still have to prove that $w$ is a viscosity solution on $]\bar{x}, +\infty[$. 
  First, for all $x \in ]\bar{x}, +\infty[, w'(x) = 1$. Moreover,
    \[
    \forall x > \bar{x}, -\mathcal{L}_{j_0}w = r(v_{j_0}(\bar{x}) + x - \bar{x}) - \mu\beta(k_{j_0}).
    \]
    So using :
    \[
    rv_{j_0}(\bar{x}) > \mu \beta(k_{j_0})
    \]
    we have that $-\mathcal{L}_{j_0}w > 0$. Finally, for all $j \in \mathbb{I}^c \backslash \{j_0\}$, we have
    \[
    \forall x > \bar{x}, w(x) >  x + a_j - \gamma |k_{j_0} - k_j| \geq v_j(x - \gamma|k_{j_0} - k_j|).
    \]
    For $i \in \mathbb{I}$, as $\bar{x} - \gamma|k_i - k_{j_0}| > b_i$, for all $x > \bar{x}$, $v'_i(x - \gamma|k_i - k_{j_0}|) = 1$. Thereafter,
    \[
    \begin{split}
      \forall x > \bar{x}, v_i(x-\gamma|k_i - k_{j_0}|) - w(x) &= v_i(\bar{x} - \gamma |k_i - k_{j_0}|) + x - \bar{x} - (v_{j_0}(\bar{x}) + x - \bar{x})\\
      &= v_i(\bar{x} - \gamma|k_i - k_{j_0}|) - v_{j_0}(\bar{x})\\
      &\leq 0.
\end{split}
    \]
    We proved that $w$ is a viscosity solution to the variational inequality so by uniqueness, $w=v_{j_0}$, which is in contradiction with $j_0 \in \mathbb{I}^c$ and the result is proved.
\end{pf}

Lemma \ref{lem_bi_finit} ensures that if $x$ is large enough, we have $v_i(x + h) = v_i(x) + h$. This property, with the left boundary condition at $\gamma k_i$ is enough to build a numerical scheme. However, we can prove more about the dividend region. Proposition \ref{prop_div_reg} below specifies the form of the dividend region under certain assumption and builds on the two next lemmas.

\begin{defi}
  We say that $x\in \Omega_i$ is a left border (resp. right border) of a subset $\mathcal{E}$ if there exists $\epsilon > 0$ and $(x_n)_{n \in \mathbb{N}} \notin \mathcal{E}$ such that
  \[
  \lim_{n \rightarrow +\infty} x_n = x
  \]
  and
  \[
  \forall y \in ]x,x+\epsilon[, y \in \mathcal{E}.
  \]
\end{defi}

\begin{lem}\label{left_border}
  For all $i\in [1,N]$, if $a_i$ is a left border of $\mathcal{D}_i$ such that $a_i \in \mathcal{S}_{ij}$ then $a_i - \gamma |k_i - k_j|$ is a left border of $\mathcal{D}_j$.
\end{lem}
\begin{pf}
  Take $a_i$ a left border of $\mathcal{D}_i$ such that $a_i \in \mathcal{S}_{ij}$. There exists $\epsilon > 0$ such that
  \[
  \forall x \in [a_i, a_i + \epsilon[, v_i(x) = v_i(a_i) + x - a_i .
  \]
  And $a_i \in \mathcal{S}_{ij}$ so :
  \[
  \forall x \in [a_i, a_i + \epsilon[, v_i(x) = v_j(a_i - \gamma |k_i - k_j|) + x - a_i.
  \]
  Since
  \[
  v_i(x) \geq v_j(x-\gamma|k_i - k_j|)
  \]
  we have :
  \[
  \forall x \in [a_i, a_i + \epsilon[, v_j(x - \gamma|k_i - k_j|) \leq v_j(a_i - \gamma|k_i - k_j|) + x - a_i
  \]
  and using Lemma \ref{sup_un}, we obtain
  \[
  \forall x \in [a_i, a_i + \epsilon[, v_j(x - \gamma |k_i - k_j|) = v_j(a_i - \gamma |k_i - k_j|) + x - a_i
  \]
  so $[a_i-\gamma |k_i - k_j|, a_i - \gamma|k_i - k_j| + \epsilon[ \subset \mathcal{D}_j$.
      Moreover, if there exists $\delta > 0$ such that $[a_i - \gamma|k_i - k_j| - \delta, a_i - \gamma|k_i - k_j|] \subset \mathcal{D}_j$ then
      \[
      \begin{split}
      \forall x \in ]a_i - \gamma|k_i - k_j|-\delta, a_i - \gamma|k_i - k_j|], v_j(x) &= v_j(a_i - \gamma|k_i - k_j|) + x - a_i + \gamma|k_i - k_j|\\
  &=v_i(a_i) + x- a_i + \gamma|k_i - k_j|\\
  &>v_i(x+\gamma|k_i - k_j|)
      \end{split}
      \]
      which is a contradiction and the result is proved.
\end{pf}

\begin{lem}\label{aipgqki}
  For all $i \in [1,N]$, if $a_i$ is a left border of $\mathcal{D}_i$ then
  \[
  \left\{
  \begin{split}
 	&a_i \in \bar{\mathcal{C}}_i \Rightarrow a_i \geq k_i \text{ and } v_i(a_i) = \frac{\mu \beta(k_i)}{r},\\
  &a_i \in \mathcal{S}^+_i \Rightarrow a_i \geq k_i \text{ and } v_i(a_i) = \frac{\mu \beta(k_l)}{r}  \text{ with } l = \min (j > i, a_i - \gamma |k_i - k_j| \notin \mathcal{S}^+_j ),\\
  &a_i \in \mathcal{S}^-_i \Rightarrow a_i < k_i \text{ and } v_i(a_i) = \frac{\mu \beta(k_l)}{r}  \text{ with } l = \max (j < i, a_i - \gamma |k_i - k_j| \notin \mathcal{S}^-_j ).
	\end{split}
	\right.   
  \]
\end{lem}

\begin{pf}
Take $a_i$ a left border of $\mathcal{D}_i$.\\
\newline
Let's prove first that
\begin{equation}\label{supsup}
v_i(a_i) \geq \frac{\mu \beta(k_i) - \alpha((k_i - a_i)^+)}{r}.
\end{equation}
As $-\mathcal{L}_iv_i \geq 0$ and $v_i'(x) = 1$ on $[a_i, a_i+\epsilon[$, we have
      \[
      \forall x \in [a_i, a_i +\epsilon[, -(\mu\beta(k_i) - \alpha((k_i - x)^+)) + r v_i(x) \geq 0.
      \]
      So,
      \[
      v_i(a_i) \geq \frac{\mu \beta(k_i) - \alpha((k_i - a_i)^+)}{r}.
      \]
\textit{First case :} $a_i \in \bar{\mathcal{C}}_i$. It exists $\delta$ such that $]a_i - \delta, a_i[ \subset \mathcal{C}_i$ and $v_i$ is $C^2$ over this interval (see Proposition \ref{derivC2}). Using Lemma \ref{sup_un}, we have $v''_i(a_i^-) \leq 0$. So using the differential equation satisfied by $v_i$ over $]a_i-\delta,a_i[$, we have
      \[
      0 \geq rv_i(a_i) - \mu\beta(k_i) + \alpha((k_i - a_i)^+).
      \]
      So
      \[
      v_i(a_i) = \frac{\mu\beta(k_i) - \alpha((k_i - a_i)^+)}{r}.
      \]
      Suppose $a_i < k_i$ then it exists $\epsilon > 0$ such that $]a_i, a_i + \epsilon[ \in \mathcal{D}_i \cap ]\gamma k_i, k_i[$. Then
      \[
      \left\{
      \begin{split}
        &v_i(x) = v_i(a_i) + x - a_i,\\
        &-\mathcal{L}_iv_i(x) \geq 0,\\
      \end{split}
      \right.
      \]
      so
      \[
      -(\mu\beta(k_i) - \alpha(k_i - x)) + r \left(\frac{\mu \beta(k_i) - \alpha(k_i - x)}{r} + x - a_i\right) \geq 0.
      \]
      It follows that
      \[
      \alpha(k_i - x) - \alpha(k_i - a_i) + r(x - a_i) \geq 0
      \]
      which is a contradiction since $\alpha' > r$.\\
\newline
      \textit{Second case} : $a_i \in \mathcal{S}^+_{ij}$. In this case, using Lemma \ref{left_border}, we know that $a_i - \gamma|k_i - k_j|$ is a left border of $\mathcal{D}_j$. Therefore, taking $l=\min\{j>i, a_i - \gamma|k_i - k_j| \notin \mathcal{S}^+_j\}$ we can use the first case and we have
      \[
      a_i - \gamma|k_i - k_l| \geq k_l
      \]
      and
      \[    
		v_i(a_i) = v_l(a_i - \gamma |k_i - k_l|)      
      \]
      which implies that
      \[
      a_i \geq k_i 
      \]
      and
      \[
      v_i(a_i) = \frac{\mu \beta(k_l)}{r}
      \]
      and the result is proved.\\
      \newline
      \textit{Third case} : $a_i \in \mathcal{S}^-_{ij}$. In this case, using Lemma \ref{left_border}, we know that $a_i - \gamma|k_i - k_j|$ is a left border of $\mathcal{D}_j$. Therefore, taking $l=\max\{j<i, a_i - \gamma|k_i - k_j| \notin \mathcal{S}^-_j\}$ we can use the first case and we have
      \[    
		v_i(a_i) = v_l(a_i - \gamma |k_i - k_l|) = \frac{\mu \beta(k_l)}{r}.      
      \]
      But remember that (\ref{supsup}) : 
      \[      
		v_i(a_i) \geq \frac{\mu \beta(k_i) - \alpha((k_i - a_i)^+)}{r}
      \]
      so
      \[      
		  \frac{\mu \beta(k_l)}{r} \geq \frac{\mu \beta(k_i) - \alpha((k_i - a_i)^+)}{r}  
      \]
      which is impossible if $a_i \geq k_i$ and the result is proved.
\end{pf}

\begin{pro}\label{prop_div_reg}
  For all $i \in [1,N]$, if $b_i \notin \mathcal{S}_i$ and $\mu\beta(k_i) > \alpha((1-\gamma)k_i)$ then $\mathcal{D}_i = [b_i,+\infty[ \cup \mathcal{E}$ where $\mathcal{E}$ is a set with empty interior.
\end{pro}
\begin{pf}
  Suppose there is another non-empty interior subset in $\mathcal{D}_i$, then it exists a right and a left border that we note $d_i$ and $g_i$. We prove the result in two steps.\\
  \newline
  \textit{First step}: Suppose $g_i = \gamma k_i$.\\
  There exists $\epsilon > 0$ such that $[\gamma k_i, \gamma k_i + \epsilon] \subset \mathcal{D}_i$. So for all $x \in [\gamma k_i, \gamma k_i + \epsilon], v(x) = x - \gamma k_i$ and
      \[
      -(\mu \beta(k_i) - \alpha(k_i - x)) + r(x - \gamma k_i) \geq 0.
      \]
      But
      \[
      \lim_{x \rightarrow \gamma k_i} -(\mu \beta(k_i) - \alpha(k_i - x)) + r(x-\gamma k_i) = -\mu \beta(k_i) + \alpha((1-\gamma) k_i) < 0
      \]
      which is a contradiction.\\
      \newline
      \textit{Second step}: $g_i > \gamma k_i$\\
      Using Lemma \ref{aipgqki} we know that $g_i \geq k_i$, so $d_i > k_i$. This means that there exists $\epsilon > 0$ such that $d_i - \epsilon \geq k_i$ and
      \[
      \forall x \in ]d_i - \epsilon, d_i], v_i(x) = v_i(d_i) + x - d_i.
  \]
  Using then that $-\mathcal{L}_iv_i(x) \geq 0$ over $]d_i -\epsilon,d_i]$, we have that
  \[
  v_i(d_i) \geq \frac{\mu \beta(k_i)}{r}.
  \]
  Using again Lemma \ref{aipgqki}, as $b_i \notin \mathcal{S}_i$, then
  \[
  v_i(b_i) = \frac{\mu \beta(k_i)}{r}
  \]
  which is a contradiction since $b_i > d_i$ and $v_i$ is a strictly increasing function.
\end{pf}

Those theoretical results (in particular proposition (\ref{sol_visc}) and Lemma (\ref{lem_bi_finit})) allow us to define the final form of the localized HJB equation which will be numerically solved in the next section :

\begin{coro}\label{sol_visc_comp}
The shareholders value functions $(v_i)_{i\in [1,N]}$ are the unique continuous viscosity solutions to the HJB variational inequality :
\begin{equation}\label{visc_eq_comp}
\begin{split}
&\forall i \in [1,N], \forall x \geq \gamma k_i,\\
&\min\left\{-\mathcal{L}_iv_i(x), v_i'(x) - 1, v_i(x) - \max_{j\neq i} v_j(x-\gamma|k_i - k_j|) \right\} = 0\\
\end{split}
\end{equation}
with boundary conditions
\begin{equation} \label{defaultboundary_comp}
\forall i \in [1,N],
\left\{ 
\begin{split}
&v_i(\gamma k_i) = 0,\\
&v'_i(b_i) = 1,\\
\end{split}
\right.
\end{equation}
where $b_i$ is defined in Lemma \ref{lem_bi_finit}.
\end{coro}

\section{Numerical Approximation}

The aim of this section is to produce a grid and a discretization by means of central, forward and backward differencing of the boundary problem (\ref{visc_eq_comp}). We prove that the scheme is monotone and therefore converges to the solution (see \cite{fl:07}). 

\subsection{Finite difference method}
First we make the following change of variable :
\[
w_i(x) = v_i(x+\gamma k_i).
\]
Clearly, $w_i$ is the unique viscosity solution of the equation 

\begin{equation}\label{eq_min}
\min\Big(- \bar{\mathcal{L}_i} w_i(x),w_i’(x)-1,w_i(x)-\max(w_{i-1}(x),w_{i+1}(x-2\gamma h))\Big)= 0
\end{equation}
with, 
\[
\bar{\mathcal{L}_i}\phi (x)= (\mu\beta(k_i) - \alpha(((1-\gamma)k_i-x)^+))\phi'(x) + \frac{\beta(k_i)^2\sigma^2}{2}\phi''(x) - r \phi
\]
 We also define the operator $\tilde{\mathcal{I}}$ for a function $\phi$ and a point $x$ that gives the linear interpolation of $\phi$ at the point $x-2\gamma h$. 

\begin{rem}
Remind that there isn't investment for $i=N$ neither disinvestent for $i=1$. So in the precedent definition, we suppose that the condition $v_i(x)-v_{i-1}(x-\gamma h)$ (resp. $v_i(x) - v_{i+1}(x-\gamma h)$) fades away when $i=1$ (resp. $i = N$).
\end{rem}
From the previous section, we know that for all $i\in[1,N]$, there exists $b_i$ such that $w'_i(x) = 1$ over $[b_i,+\infty[$. We don't know a priori the boundary $b_i$ but taking $x_{max}$ big enough we will have $x_{max} > \max_i\{b_i\}$ and therefore the boundaries conditions :
\[
\forall i \in [1,N], w_i'(x_{max}) = 1.
\]
Problem (\ref{eq_min}) is therefore solved on the computational domain
\begin{equation}
(x,k) \in [0,x_{max}] \times [k_1,k_N]
\end{equation}
with the regular grid $(x_l)_{l\in[1,M]} = \{(l-1) \Delta x\}_{l \in [1,M]}$ with 
\[
\Delta x = \frac{x_{max}}{M-1}
\]
in order to have $x_M = x_{max}$. 
Let $W_{l,i}$ be the approximate solution of equation (\ref{eq_min}) at $(x_l,k_i)$ for every $i\in [1,N]$ and $l\in [1,M]$. We use a direct method similar to \cite{hfl:12a} to discretize equation (\ref{eq_min}) as well as central differencing as much as possible in order to improve the efficiency.
More precisely, we solve, setting $\tilde{\mathcal{L}}_i$ the discretization of  $\bar{\mathcal{L}_i}$,
\begin{equation}\label{eq_lin}
\begin{split}
&\rho_{l,i}\left[\theta_{l,i}\left(-\psi_{l,i}(\tilde{\mathcal{L}}W)_{l,i} + (1-\psi_{l,i})\left(\frac{W_{l,i} - W_{l-1,i}}{\Delta x} - 1\right)\right)\right]\\
=&- \rho_{l,i}(1-\theta_{l,i})(W_{l,i} - W_{l,i-1}) - (1-\rho_{l,i})(W_{l,i} - \tilde{\mathcal{I}} W_{l,i+1})
\end{split}
\end{equation}
with,
\[
\{\rho_{l,i},\theta_{l,i},\psi_{l,i}\} \in \{0,1\}
\]
Solving (\ref{eq_lin}), we make sure that at least one of the terms in \eqref{eq_min} is equal to zero. To make sure to satisfy that all the terms are positive, we choose $(\rho, \theta, \psi)$ such that
\begin{equation}\label{discret}
\begin{split}
\{\rho_{l,i},\theta_{l,i},\psi_{l,i}\} = argmin_{\substack{\tilde \rho \in \{0,1\}\\ \tilde\theta \in \{0,1\}\\ \tilde\psi \in \{0,1\}}}\Big\{&\tilde \rho_{l,i}\left[\tilde \theta_{l,i}\left(-\tilde \psi_{l,i}(\tilde{\mathcal{L}}W)_{l,i} + (1-\tilde\psi_{l,i})\left(\frac{W_{l,i} - W_{l-1,i}}{\Delta x} - 1\right)\right)\right]\\
&+ \tilde\rho_{l,i}(1-\tilde\theta_{l,i})(W_{l,i} - W_{l,i-1}) + (1-\tilde\rho_{l,i})(W_{l,i} - \tilde{\mathcal{I}} W_{l,i+1})\Big\}.
\end{split}
\end{equation}
This ensures that all the terms in the equation (\ref{eq_min}) are positives which implies that (\ref{eq_lin}) with (\ref{discret}) is the rightful dicretization of (\ref{eq_min}).
The terminal boundary condition $w_i'(x_M) = 1$ is classically discretized :
\[
W_{M,i} = W_{M-1,i} + \Delta x, i \in [1,N]
\]
and we also have
\[
W_{0,i} = 0, i \in [1,N].
\]

\begin{rem}
A penalty method can be used to solve that kind of equation. However, to avoid the calibration of the penalty parameter we prefer the direct approach (see \cite{hfl:12c}) .
\end{rem}

If we denote
\[
\left\{
\begin{split}
&C_1(x_l,k_i) = \mu\beta(k_i) - \alpha(((1 - \gamma)k_i - x_l)^+),\\
&C_2(x_l,k_i) = \frac{\sigma^2\beta(k_i)^2}{2} > 0,\\
\end{split}
\right.
\]
then to satisfy the positive coefficient condition and to maximize the efficiency, the discretized operator $\tilde{\mathcal{L}}_i$ is given by :
\[
(\tilde{\mathcal{L}}W)_{l,i} =
\left\{
\begin{split}
&C_2(x_l,k_i)\frac{W_{l+1,i} + W_{l-1,i} - 2 W_{l,i}}{\Delta x^2} + C_1(x_l,k_i)\frac{W_{l+1,i} - W_{l-1,i}}{2 \Delta x} - r W_{l,i}\\
&\text{ if } 2C_2(x_l,k_i) \geq |C_1(x_l,k_i)|\Delta x \text{ (central differencing)}.\\
&C_2(x_l,k_i)\frac{W_{l+1,i} + W_{l-1,i} - 2 W_{l,i}}{\Delta x^2} + C_1(x_l,k_i)\frac{W_{l+1,i} - W_{l,i}}{\Delta x} - r W_{l,i}\\
  &\text{ if } 2C_2(x_l,k_i) < |C_1(x_l,k_i)|\Delta x \text{ and } C_1(x_l,k_i) \geq 0 \text{ (forward differencing)}.\\
  &C_2(x_l,k_i)\frac{W_{l+1,i} + W_{l-1,i} - 2 W_{l,i}}{\Delta x^2} + C_1(x_l,k_i)\frac{W_{l,i} - W_{l-1,i}}{\Delta x} - r W_{l,i}\\
  &\text{ if } 2C_2(x_l,k_i) < |C_1(x_l,k_i)|\Delta x \text{ and } C_1(x_l,k_i) < 0 \text{ (backward differencing)}.\\
\end{split}
\right.
\]

\begin{pro}
The scheme is monotone, consistent and stable.
\end{pro}
\begin{pf}
  The scheme is, as a finite difference scheme, consistent. Moreover, we check easily that in $-(\mathcal{L}W)_{i,l}$, the coefficients in front of $W_{l-1,i}, W_{l+1,i}, W_{l,i-1}$ are negatives. So are the coefficients in front of $W_{k,i+1}$ for $k$ acting in the interpolation $\mathcal{\tilde I}W_{l,i+1}$. On the contrary, the coefficient in front of $W_{l,i}$ is positive which proves the monotony. We still have to prove the stability i.e. to prove that for all $\Delta x$, the schema has a solution $(W_{l,i})_{i,l}$ which is uniformly bounded independently of $\Delta x$. First, equation (\ref{eq_lin}) implies that
  \begin{equation}\label{14}
  \forall i \in [1,N], \forall l \in [2,M], W_{l,i} \geq W_{l-1,i} + \Delta x \geq W_{l-1,i}
  \end{equation}
  so the sequence $l \rightarrow W_{l,i}$ is increasing. Let's prove that $W_{M,i}$ is bounded independently of $\Delta x$. We know by the terminal boundary condition that
  \[
  \forall i \in [1,N], W_{M,i} = W_{M-1,i} + \Delta x.
  \]
  Let's note $d=\max\{j \in [1,M], W_{j,i} > W_{j-1,i} + \Delta x\}$. By equation (\ref{eq_lin}), we have one of the three next assertions which is true
  \begin{enumerate}
  \item $-(\tilde{\mathcal{L}}W)_{d,i} = 0$.
  \item $W_{d,i} - W_{d,i-1} = 0$.
  \item $W_{d,i} - \tilde{\mathcal{I}}W_{d,i+1} = 0$.  
  \end{enumerate}
  and by definition of $d$ :
  \begin{equation}\label {15}
  W_{d+1,i} = W_{d,i} + \Delta x.
  \end{equation}
  \textit{Case 1 :} Using the discretized operator, we have in the central differencing case :
  \[
  -C_2(x_d,k_i)\frac{W_{d+1,i} + W_{d-1,i} - 2 W_{d,i}}{\Delta x^2} - C_1(x_d,k_i)\frac{W_{d+1,i} - W_{d-1,i}}{2 \Delta x} + r W_{d,i} = 0.
  \]
  Then, using (\ref{15}),
  \[
  -C_2(x_d,k_i)\frac{W_{d-1,i} - W_{d,i} + \Delta x}{\Delta x^2} - C_1(x_d,k_i)\frac{W_{d,i} - W_{d-1,i} + \Delta x}{2 \Delta x} + r W_{d,i} = 0.
  \]
  Factoring,
  \[
  rW_{d,i} = C_1(x_d,k_i) + \left(-\frac{C_2(x_d,k_i)}{\Delta x^2} + \frac{C_1(x_d,k_i)}{2\Delta x} \right)(W_{d,i} - W_{d-1,i} - \Delta x)
  \]
  so, using that $W_{d,i} \geq W_{d-1,i} + \Delta x$ and the central differencing inequation, we have
  \[
  W_{d,i} \leq \frac{C_1(x_d,k_i)}{r}.
  \]
  Moreover, $C_1(x_d, k_i)$ is bounded independently of $(x_d,k_i)$ by $\mu \bar{\beta}$. Then,
  \[
  W_{d,i} \leq \frac{\mu \bar{\beta}}{r}.
  \]
  But, by definition of $d$,
  \[
  W_{M,i} = W_{d,i} + (M - d) \Delta x,
  \]
  so
  \[
  W_{M,i} \leq \frac{\mu \bar{\beta}}{r} + (M - d) \Delta x \leq \frac{\mu \bar{\beta}}{r} + x_M.
  \]
  The proof for the forward and backward differencing is similar and therefore omitted.\\
\newline
\textit{Case 2 : } In this case, let's define $p = \max\{j \in [1,i-1], W_{d,j} - W_{d,j-1}> 0\}$. At this point, we necessarily have $-\tilde{\mathcal{L}}W_{d,p} = 0$. Let's prove that we also have
\[
W_{d+1,p} = W_{d,p} + \Delta x.
\]
Suppose that $W_{d+1,p} > W_{d,p} + \Delta x$. Then, using that $W_{d,p} = W_{d,i}$
\[
W_{d+1,p} >  W_{d,i}  + \Delta x.
\]
But by definition of $d$, $W_{d,i} + \Delta x = W_{d+1,i}$, so
\[
W_{d+1,p} >  W_{d+1,i}
\]
which is a contradiction since $p < i$. So we have
\[
W_{d+1,p} = W_{d,p} + \Delta x
\]
and
\[
-\tilde{\mathcal{L}}W_{d,p} = 0
\]
and we can use the first case to prove that
\[
W_{d,i} \leq \frac{\mu \bar{\beta}}{r} + x_M.
\]
The proof in the third case is similar and is therefore omitted.\\
Finally, we have proved in all cases that
\[
\forall i \in [1,N], \forall l \in [1,M], W_{l,i} \leq \frac{\mu \bar{\beta}}{r} + x_M
\]
which is a bound independent of $\Delta x$ and the result is proved.

\end{pf}

\subsection{Matrix Form of the Discretized Equations}
We denote $U$ the vector of size $N\times M$ and $\hat{U}_i$ for $i\in[1,N]$ the vectors of size $M$ such that
\[
\hat{U}_i = (W_{l,i})_{l \in [1,M]}, \qquad U = (\hat{U}_1, \hat{U}_2,...,\hat{U}_N).
\]
With
\[
j = l + (i-1) M
\]
we have
\[
\forall j \in [1,N \times M], U_j = W_{l,i}.
\]
Then we can write (\ref{eq_lin}) in a linear matrix form as follows :
\begin{equation}\label{matrix_form}
A(\rho,\psi,\theta) U + B(\rho,\psi,\theta) = 0
\end{equation}
where $A(\rho,\psi,\theta)$ is a square tridiagonal block matrix of size $N \times M$ :\\
\begin{tikzpicture}
\matrix(M)[matrix of math nodes, left delimiter=(, right delimiter=),nodes={minimum size=1cm,scale=1.2}, above delimiter = \{]{
Z^1&D^1&&&&&\\
C^2&&&&&&\\
&\hphantom{\hspace{1cm}}&\hphantom{\hspace{1cm}}&\hphantom{\hspace{1cm}}&\hphantom{\hspace{1cm}}&\hphantom{\hspace{1cm}}&\\
&&C^i&Z^i&D^i&&\\
&\hphantom{\hspace{1cm}}&&&&&\\
&&&&&&D^{N-1}\\
&&&&&C^N&Z^N\\
};
\tikzset{pointille/.style={loosely dotted,thick}};
\draw[pointille] (M-1-1.south east)--(M-4-4.north west);
\draw[pointille] (M-2-1.south east)--(M-4-3.north west);
\draw[pointille] (M-1-2.south east)--(M-4-5.north west);
\draw[pointille] (M-4-4.south east)--(M-7-7.north west);
\draw[pointille] (M-4-3.south east)--(M-7-6.north west);
\draw[pointille] (M-4-5.south east)--(M-6-7.north west);
\node[above=2mm] at (M.north) {$N \times M$};
\node[scale=1.8,below=2.8cm] at (M-1-2.south){$0$};
\node[scale=1.8,above=2.8cm] at (M-7-6.north){$0$};
\end{tikzpicture}\\
with $Z^i$ the tridiagonal matrix of size $M$ given by,
\[
Z^i = T^i + (\rho_{l,i}(1-\theta_{l,i}) + (1 - \rho_{l,i})) Id
\]
with the coefficients of $T^i$ given, for all $l \in [2, M-1]$, by :
\[
\left\{
\begin{split}
&t_{l,l}^i =  \rho_{l,i}\theta_{l,i}\left(\psi_{l,i} \left(r + \frac{2C_2(x_l,k_i)}{\Delta x^2} + \frac{|C_1(x_l,k_i)|}{\Delta x} 1_{\{2C_2(x_l,k_i) < \Delta x|C_1(x_l,k_i)| \}}\right) + \frac{(1 - \psi_{l,i})}{\Delta x}\right), \\
&t_{l,l+1}^i =  \rho_{l,i}\theta_{l,i} \left(- \psi_{l,i} \left( \frac{C_2(x_l,k_i)}{\Delta x^2} + \frac{C_1(x_l,k_i)}{2 \Delta x} + \frac{|C_1(x_l,k_i)|}{2\Delta x} 1_{\{2C_2(x_l,k_i) < \Delta x|C_1(x_l,k_i)| \}}\right)\right) ,\\
&t_{l,l-1}^i =  \rho_{l,i}\theta_{l,i} \left(- \psi_{l,i} \left( \frac{C_2(x_l,k_i)}{\Delta x^2} - \frac{C_1(x_l,k_i)}{2 \Delta x} + \frac{|C_1(x_l,k_i)|}{2\Delta x} 1_{\{2C_2(x_l,k_i) < \Delta x|C_1(x_l,k_i)| \}}\right)-\frac{(1-\psi_{l,i})}{\Delta x}\right).\\
\end{split}
\right.
\]
To fulfill the Dirichlet conditions, we force :
\[
\forall i \in [1,N], \forall l \in [1,M], 
\left\{
\begin{split}
&Z_{1,l}^i = \delta_{1l},\\
&D_{1,l}^i = C^i_{1,l} = 0.\\
\end{split}
\right.
\]
Also to satisfy the right boundary conditions, the controls are fixed for $l=M$ : 
\[
\forall i \in [1,N], \left\{
\begin{split}
\rho_{M,i} &= 1,\\
\theta_{M,i} &= 1,\\
\psi_{M,i} & = 0.
\end{split}
\right.
\]
By doing this we make sure that at the right boundary of the grid, we are in the dividend region.\\
$C^i$ is the diagonal matrix of size $M$ with
\[
\forall i \in [1,N], \forall l \in [1,M], c^i_{l,l} = -(1-\theta_{l,i})\rho_{l,i}.
\]
The form of the matrix $D^i$ is given by the operator $\tilde{\mathcal{I}}$. To have a monotone scheme, we need to use a linear operator. With a constant step, the form of the matrix is as follows :\\ 
\begin{tikzpicture}
\matrix(M)[matrix of math nodes, left delimiter=(, right delimiter=),nodes={minimum size=2.8cm,scale=0.6}]{
0&&&&&&0\\
0&&&&&&0\\
-(1-\rho_{3,i})\lambda&   -(1-\rho_{3,i})(1-\lambda)&0&&&&\\
&&&\hphantom{\hspace{1.5cm}}&&&\\
&&&\hphantom{\hspace{1.5cm}}&&&\\
&&&\hphantom{\hspace{1.5cm}}&&&\\
&&&&-(1-\rho_{M,i})\lambda&-(1-\rho_{M,i})(1-\lambda)&0\\
};
\tikzset{pointille/.style={loosely dotted,thick}};
\draw[pointille] (M-1-1.east)--(M-1-7.west);
\draw[pointille] (M-2-1.east)--(M-2-7.west);
\draw[pointille] (M-3-3.south east)--(M-7-7.north west);
\draw[pointille] (M-3-1.south east)--(M-7-5.north west);
\draw[pointille] (M-3-2.south east)--(M-7-6.north west);
\draw[pointille] (M-1-1.south)--(M-2-1.north);
\draw[pointille] (M-1-7.south)--(M-2-7.north);
\node[scale=1.5,below=1.5cm] at (M-3-2.south){$0$};
\node[scale=1.5,above=1.5cm] at (M-7-6.north){$0$};
\node[scale=1.3,left=10.3cm] at (M.east){$D^i = $};
\end{tikzpicture}\\
The offset $d$ with respect to the diagonal in the matrix $D^i$, i.e þÀthe number of zero lines at the beginning of $D^i$, is equal to
\[
d = 1 + E\left(\frac{2\gamma h}{\Delta x}\right)
\]
where $E$ is the floor function. We also define the fractional part :
\[
\lambda = \frac{2\gamma h}{\Delta x} - E\left(\frac{2\gamma h}{\Delta x}\right).
\]
Last, the vector $B$, is of size $N \times M$ and satisfies
\[
\forall j \in [1,N\times M], b_j = -\rho_{l,i}\theta_{l,i}(1-\psi_{l,i}), \qquad j = l + (i-1)M.
\]

\begin{defi}
A matrix $A$ is a M-matrix if $A$ is non singular, the off-diagonal coefficients of $A$ are negatives and $A^{-1} \geq 0$.
\end{defi}

\begin{rem}
A matrix such that $A^{-1} \geq 0$ satisfies
\[
X \geq 0 \Leftrightarrow A^{-1}X \geq 0.
\]
\end{rem}

\begin{lem}\label{AiMmatrice}
The matrices $Z^i(\rho,\psi,\theta) = (z^i_{l,j})_{l,j}$ are M-matrices.
\end{lem}

\begin{pf}
Take $i\in[1,N]$. We observe that the matrix $Z^i$ satisfies for all $l$, $z^i_{ll} > 0$ and for all $l \neq j$, $z^i_{lj} \leq 0$. Moreover $Z^i$ is a diagonally dominant matrix. Indeed,
\[
\forall  l \in [2,M], z^i_{ll} - \sum_{j\neq l} |z^i_{l,j}| =
\left\{
\begin{split}
&r, \qquad \psi_{li}\rho_{l,i} \theta_{l,i} = 1,\\
&0, \qquad \psi_{li} = 0 \text{ et } \rho_{l,i} \theta_{l,i} = 1, \\
&1, \qquad \rho_{li} \theta_{li} = 0.\\
\end{split}
\right.
\]
For $l=1$, Dirichlet condition dictates that
\[
z^i_{11} - \sum_{j\neq 1} |z^i_{1,j}| = 1.
\]
However the matrix is not a strict diagonally dominant matrix since when there is distribution of dividends (i.e. $\psi_{li} = 0 \text{ and } \rho_{l,i} \theta_{l,i} = 1$), the sum of the line coefficients is equal to zero and thus the classical technique used in \cite{hfl:12a} does not apply. Another way to prove that $Z^i(\rho,\psi,\theta)$ is a M-matrix is to find a M-size vector $W$ such that $W > 0$ and $Z^iW > 0$ (see \cite{pb:74}).
Let's prove that $W$ the M-size vector given by
\[
\forall l \in [1,M], W_l = 1 + l\epsilon
\]
with
\[
\epsilon = \frac{r\Delta x}{\mu \bar{\beta}} > 0
\]
satisfies this condition for all $i \in [1,N]$.
First, as $Z^i$ is a tridiagonal matrix, we have
\[
\forall l \in [2,M-1], (Z^iW)_l = z^i_{l,l-1}(1 + (l-1)\epsilon) + z^i_{l,l}(1 + l\epsilon) + z^i_{l,l+1}(1+(l+1)\epsilon).
\]
For $l=1$
\[
(Z^iW)_1 = (1 + \epsilon) 
\]
and for $l=M$
\[
(Z^iW)_M = \frac{1 + M \epsilon}{\Delta x} - \frac{1 + (M -1)\epsilon}{\Delta x}  .
\]
Then, for all $l \in [2,M]$, we have :
\[
(Z^iW)_l =
\left\{
\begin{aligned}
&(1+(l-1)\epsilon)r + \epsilon (z^i_{ll}+2z^i_{l,l+1}), &\rho_{li} \theta_{li}\psi_{li} = 1,\\
&\frac{\epsilon}{\Delta x}, &\rho_{li}\theta_{li} = 1 \text{ et } \psi_{li} = 0,\\
&1 + l\epsilon, &\rho_{li}\theta_{li} = 0.\\
\end{aligned}
\right.
\]
Moreover, when $\rho_{li} \theta_{li}\psi_{li} = 1$, we have
\[
z^i_{ll} + 2 z^i_{l,l+1} = r - \frac{C_1(x_l,k_i)}{\Delta x}.
\]
So,
\[
\begin{split}
(1+(l-1)\epsilon)r + \epsilon (z^i_{ll}+2z^i_{l+1,l}) &= (l-1)r\epsilon + r + \epsilon \left(r - \frac{C_1(x_l,k_i)}{\Delta x} \right)\\
&\geq lr\epsilon + r - \epsilon \frac{C_1(x_l,k_i)}{\Delta x}.\\
\end{split}
\]
Using the definition of $\epsilon$ and the fact that
\[
\forall i \in [1,N], \forall l \in [1,M], C_1(x_l,k_i) \leq \mu \bar{\beta} 
\]
we have that
\[
(1+(l-1)\epsilon)r + \epsilon (z^i_{ll}+2z^i_{l+1,l}) \geq lr\epsilon.
\]
So
\[
\forall l \in [1,M], (Z^iW)_l > 0.
\]
And we conclude that $Z^i$ is a M-matrix. 
\end{pf}

\begin{coro}\label{cor1}
The matrix $A(\rho,\theta, \psi)$ is a M-matrix.
\end{coro}
\begin{pf}
To prove the result we use the same technique as in Lemma \ref{AiMmatrice}. Choose
\[
\epsilon = \frac{r\Delta x}{\mu \bar{\beta}} > 0
\] 
and
\[
0 < \eta < (d+\lambda)\epsilon.
\]
Thus, we define $W$ the $N \times M$-size vector as follows
\[
\forall i \in [1,N], \forall l \in [1,M], W_{l + (i-1) M} = 1 + l\epsilon + i\eta.
\]
Then, using the results of Lemma \ref{AiMmatrice}, we obtain that for all $i \in [1,N]$ and for all $l \in [2,M-1]$,
\[
\rho_{li}\theta_{li} = 1 \Rightarrow (AW)_{l + (i-1)M} \geq \min\left(\frac{\epsilon}{\Delta x}, lr\epsilon\right) + (z^i_{l,l} + z^i_{l,l-1} + z^i_{l,l+1}) i\eta .
\]
Using that the matrices $Z^i$ are diagonally dominant, we have that 
\[
\rho_{li}\theta_{li} = 1 \Rightarrow (AW)_{l + (i-1)M} > 0.
\]
In the case $(\theta_{li}, \rho_{l,i}) = (0,1)$, we have 
\[
(AW)_{l + (i-1)M} = 1 + l\epsilon + i\eta - ( 1 + l\epsilon + (i-1)\eta ) = \eta > 0
\]
and in the case $\rho_{l,i} = 0$, we have
\[
\begin{split}
(AW)_{l + (i-1)M} =& 1 + l\epsilon + i\eta - (1 - \lambda)[1 + (l-d)\epsilon + (i+1)\eta]\\
 &- \lambda[1 + (l-1-d)\epsilon + (i+1)\eta]\\
=& (d + \lambda)\epsilon - \eta\\
>& 0.
\end{split}
\]
Thus, we display a vector $W > 0$ such that, for all $i \in [1,N]$, for all $l\in [1,M]$,  $(AW)_{l + (i-1)M} > 0$ which proves he result.

\end{pf}

\begin{rem}
A referee pointed out that there is alternate way of proving that $A(\rho,\theta, \psi)$ is a M-matrix by using the notion of weakly-chained diagonally dominant matrices introduced in \cite{sc:74}. 
\end{rem}

\section{Convergence of the scheme}

The main result of this section is that $U$ is the solution of the following Policy iteration : 
\begin{equation}\label{pol_ite}
A^qU^{q+1} + B^q = 0
\end{equation}
with
\[
\begin{split}
A^q &= A(\rho^q, \theta^q, \psi^q),\\
B^q &= B(\rho^q, \theta^q, \psi^q), 
\end{split}
\]
\begin{algorithm}[h]
\caption{Policy Iteration}\label{algo1}
\label{inner}
\begin{algorithmic}
\STATE  $(\rho^0,\theta^0,\psi^0) = (1,1,1)$
\STATE $q=0$
\STATE $W^0 = 0$
\WHILE{Error $> \epsilon$}
\STATE Solve $W^{q+1}$ solution of 
\begin{eqnarray*}
 \begin{split}
&\rho^q_{l,i}\left[\theta^q_{l,i}\left(-\psi^q_{l,i}(\tilde{\mathcal{L}}W^{q+1})_{l,i} + (1-\psi^q_{l,i})\left(\frac{W^{q+1}_{l,i} - W^{q+1}_{l-1,i}}{\Delta x} - 1\right)\right)\right]\\
=&- \rho^q_{l,i}(1-\theta^q_{l,i})(W^{q+1}_{l,i} - W^{q+1}_{l,i-1}) - (1-\rho^q_{l,i})(W^{q+1}_{l,i} - \tilde{\mathcal{I}}(W^{q+1}_{i+1}))
\end{split}
\end{eqnarray*}
\STATE $(\rho^{q+1},\theta^{q+1},\psi^{q+1})$ solution of
\begin{eqnarray*}
  (\rho^{q+1}, \theta^{q+1},\psi^{q+1}) &= &  argmin_{(\rho,\theta,\psi) \in \{0,1\}} \Big\{  \rho_{l,i}\theta_{l,i}\Big[-\psi_{l,i}(\tilde{\mathcal{L}}W^{q+1})_{l,i} \\
& & + (1-\psi_{l,i})\left(\frac{W^{q+1}_{l,i} - W^{q+1}_{l-1,i}}{\Delta x} - 1\right)\Big]\\
& & + \rho_{l,i}(1-\theta_{l,i})(W^{q+1}_{l,i} - W^{q+1}_{l,i-1}) + (1-\rho_{l,i})(W^{q+1}_{l,i} - \tilde{\mathcal{I}}(W^{q+1}_{i+1})) \Big\}
\end{eqnarray*}
\STATE
Error$ = || W^{q+1} - W^{q}||_\infty$
     \STATE $q=q+1$
\ENDWHILE

\end{algorithmic}
\end{algorithm}
and is resumed in Theorem \ref{conv}.

\begin{theo}\label{conv}
Under the conditions
\begin{enumerate}[i)]
\item The matrices $A^q$ are M-matrix
\item $\lVert (A^q)^{-1}\rVert$ and $\lVert B^q\rVert$ are bounded regardless of $q$.
\end{enumerate}
 the scheme in Algorithm \ref{algo1} converges to the unique solution of equation (\ref{matrix_form}).
\end{theo}
\begin{pf}
The proof is classic and is posponed in appendix.
\end{pf}

\begin{lem}\label{matrice_bornee}
  The sequence $(U^q)_{q \geq 0}$ is bounded.
\end{lem}
\begin{pf}
  To prove the result, we have to prove that the matrices $(A^q)^{-1}$ and $B^q$ are bounded, regardless of $q$. First, by definition, we have that for all $q \geq 0$ and for all $j \in [1,N \times M]$, $b^q_j \leq 1$, so the matrix $B^q$ is bounded. Moreover, since $\rho^q$, $\theta^q$ and $\psi^q$ take discrete values :
\[
\forall q \geq 0, \forall l \in [1,M], \forall i \in [1,N], \rho^q_{l,i}, \theta^q_{l,i}, \psi^q_{l,i} = 0 \text{ or } 1,
\]
we have a finite number of invertible matrices $A^q$ so also a finite number of $(A^q)^{-1}$ and taking the maximum over all the possible combinations lead to a supremum of $(A^q)^{-1}$ regardless of $q$ and the result is poved.
\end{pf}

\section{Numerical results}

\subsection{Description of the optimal regions}

\begin{figure}[H]
\begin{center}
\includegraphics[width = 0.8 \textwidth]{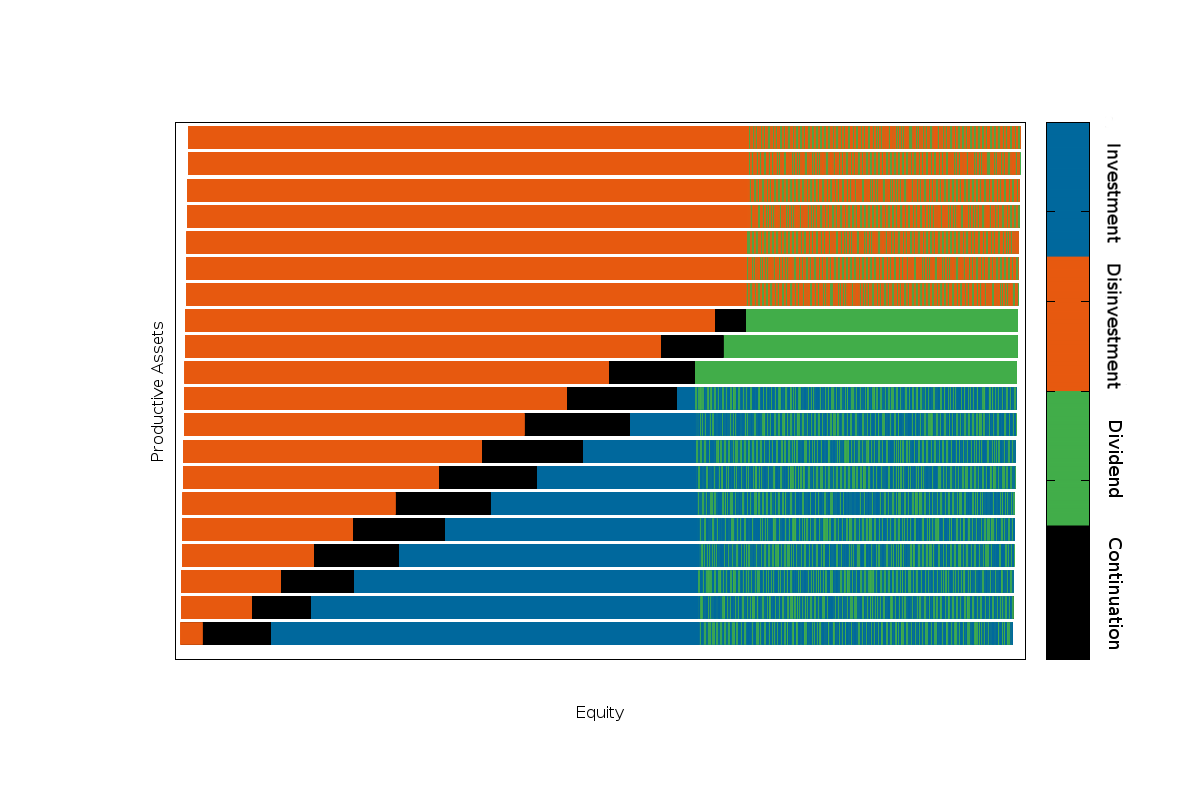}
\caption{Optimal control regions.}\label{opt_reg1}
\end{center}
\end{figure}

In chapter \ref{div_reg}, we proved that for all $i \in [1,N]$ there exists $b_i > k_i$ such that the dividend region contains at least $[b_i,+\infty[$, which allows us to define in the numerical scheme a border condition. The numerical results give us much more information about the optimal control regions. Next Proposition resumes those results :
    \begin{pro}\label{num_result}
      The optimal control regions satisfy
      \begin{enumerate}
      \item $\forall i \in [1,N], \mathcal{D}_i = [b_i,+\infty[$.
        \item $\forall i \in [2,N], \exists d_i \in \Omega_i, \mathcal{S}_i^- = [\gamma k_i, d_i]$.
          \item $\exists k^* \in [1,N], \forall i \geq k^*, \mathcal{S}^+_i = \emptyset$.
            \item $\forall i < k^*, \exists a_i \in ]d_i, b_i], \mathcal{S}^+_i = [a_i, +\infty[$.   
      \end{enumerate}
    \end{pro}
    Figure \ref{opt_reg1} illustrates Proposition \ref{num_result}. The results are obtained using a linear debt and an exponential gain function :
    \begin{itemize}
    \item Linear debt : $\alpha(x) = \lambda x$.
    \item Exponential gain function  : $\beta(x) = \bar{\beta}\left(1-\exp({-\frac{\eta}{\bar{\beta}}x})\right)$.
    \end{itemize}
    and the next values for the different parameters :
    \[
    [\mu = 0.25, \sigma = 0.40, r = 0.02, \lambda = 0.10, \bar{\beta} = 2, \eta = 1, \gamma = 1e^{-3}, N = 20, M = 1e^5 ]
    \]
With those parameters, it's enough to choose $x_{\max} = 10$ in order to have $x_{\max} > \max_i \{b_i\}$. We also choose $k_{\max} = 10$ in all the following numerical results. 

    In figure \ref{opt_reg1}, we differentiate six areas :
    \begin{enumerate}
    \item $\mathcal{S}_i^-$ (Orange zone) : Disinvestment area where the book value of equity is low in comparison to the level of the firm's productive assets. In this zone, it is optimal to disinvest to lower the risk of bankruptcy.
    \item $\mathcal{S}_i^+$ (Blue area) : Investment area where the ratio book value of equity over firm's productive assets is high. It is optimal to invest to increase the rentability via the gain function. The risk increases proportionnaly but the cash reserves protect the company against bankrupcy.
    \item $\mathcal{C}_i$ (Black area) : In between, there is the continuation area where it is optimal to not activate the controls.
    \end{enumerate}
    As proved in chapter \ref{div_reg}, we observe that on the right side of figure \ref{opt_reg1}, corresponding to a high level of equity, it is always optimal to pay dividends leading to three different areas :
    \begin{enumerate}
      \setcounter{enumi}{3}
    \item $\mathcal{S}_i^+ \cap \mathcal{D}_i$ (Green and blue area) : for a low level of productive assets. In this zone, it is optimal to pay dividends and invest until reach the optimal level $k^*$.
    \item $\mathcal{S}_i^- \cap \mathcal{D}_i$ (Green and orange area) : for a high level of productive assets. In this zone, it is optimal to disinvest until a maximum level of productive assets $k_{\max}$ in order to distribute dividends.
      \item $\mathcal{D}_i$ (Green area) : in between, it isn't optimal to invest neither to disinvest but just to pay dividends. 
    \end{enumerate}
    Those results are consistent with the economic theory. Furthermore, they bring to light two meaningful conclusions :
    \begin{itemize}
    \item It is optimal to pay dividends only once the company has reached a optimal size depending of its sector.
      \item There exists a maximum size that the company shouldn't exceed.
    \end{itemize}
    Those conclusions are directly attributable to the characteristics of the gain function chosen. Indeed, the rentability increases with the gain function which is concave with a finite limit at the infinity. So at some point, the marginal gain is small compared to the value for the shareholders to receive dividends.

    \subsection{Impact of the cost of investment}
\begin{figure}[H]
    \centering
    \begin{subfigure}[b]{0.32\textwidth}
      \includegraphics[width=\textwidth]{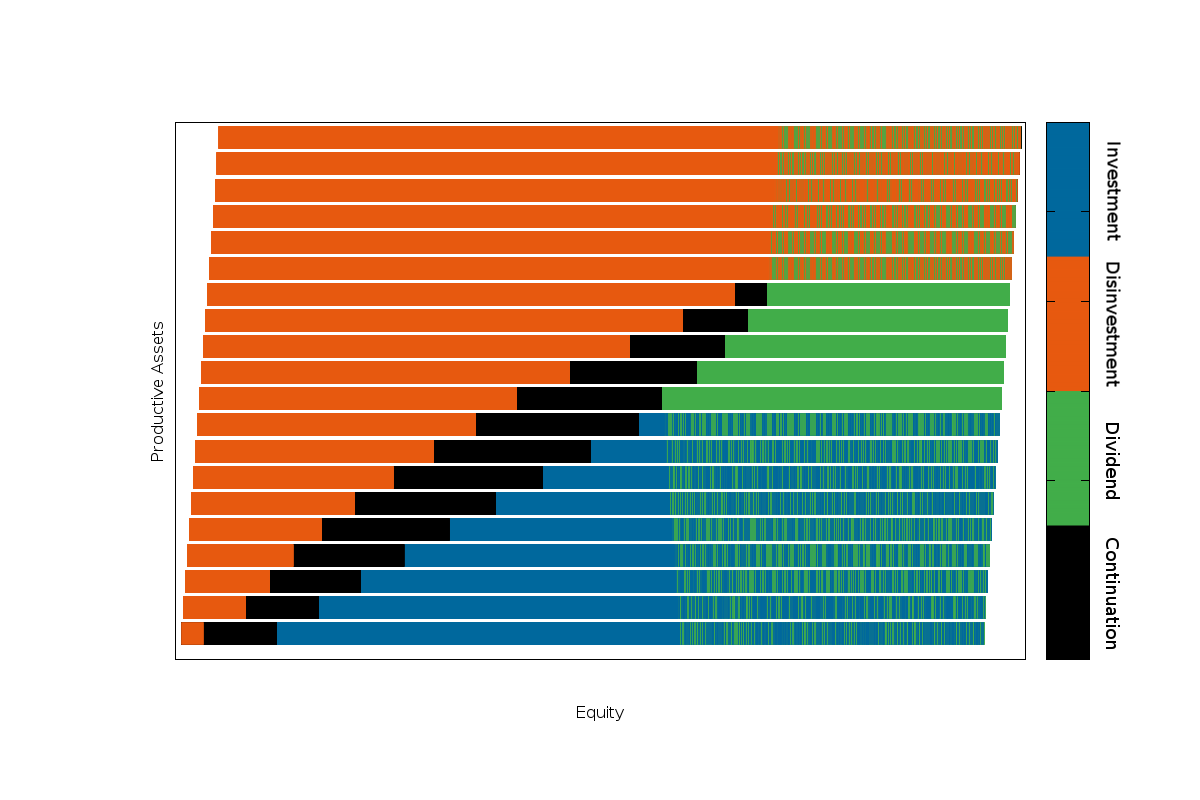}
      \caption{ $\gamma = 0.05$.}
   \end{subfigure}
   \begin{subfigure}[b]{0.32\textwidth}
     \includegraphics[width=\textwidth]{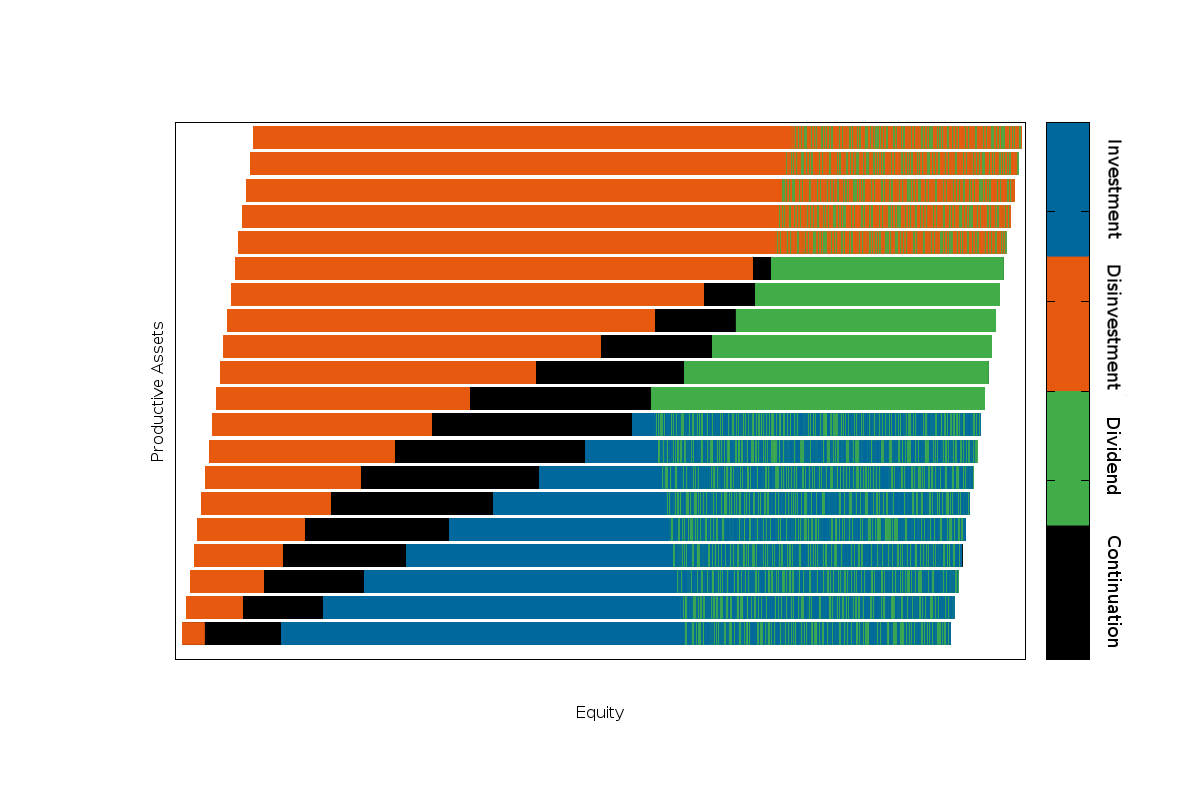}
    \caption{ $\gamma = 0.1$.}
   \end{subfigure}
  \begin{subfigure}[b]{0.32\textwidth}
    \includegraphics[width = \textwidth]{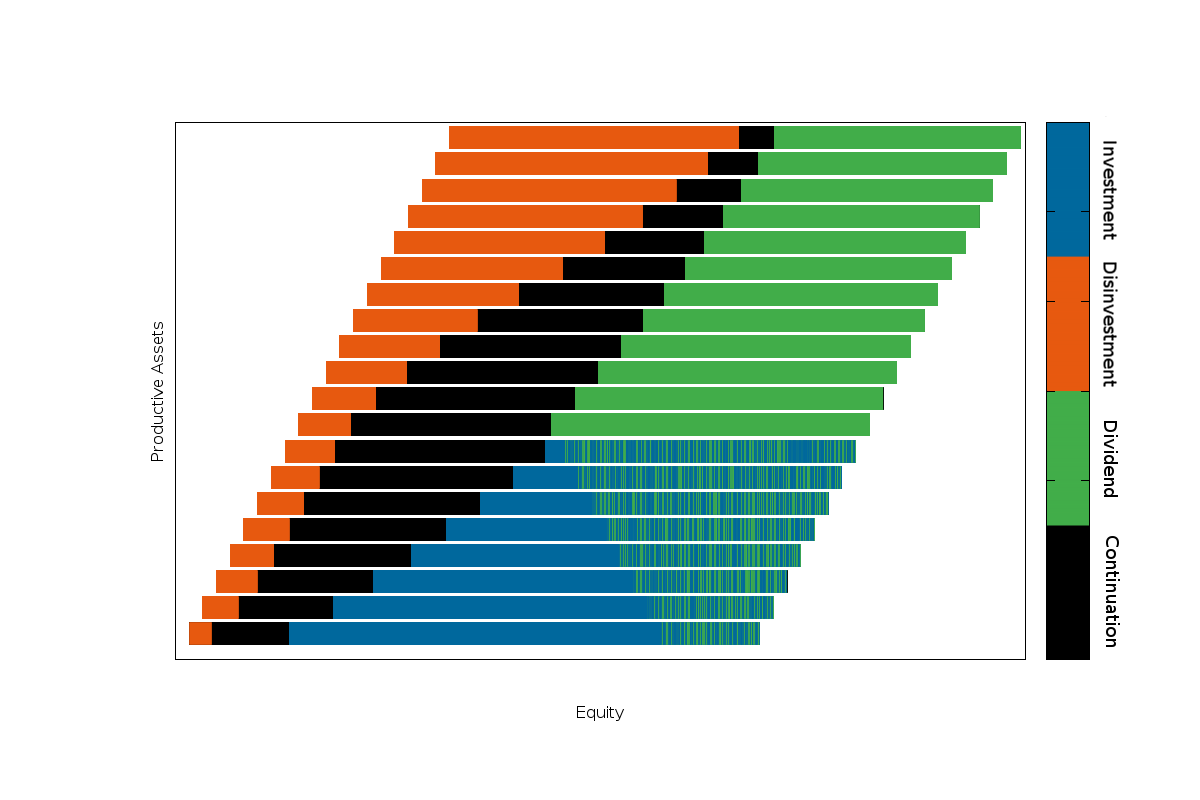}
    \caption{ $\gamma = 0.5$.}
   \end{subfigure}
\caption{ Optimal control regions for some $\gamma$ \label{opt_reg2}}

\end{figure}

    The cost of investment $\gamma$ (and of disinvestment) plays an important role in the form of the switching regions. Figure \ref{opt_reg2}  presents the optimal control regions for different values of $\gamma$ and the same parameters than figure \ref{opt_reg1}.

    We observe that the higher $\gamma$ is, the wider the continuation region is. Which is consistent since if the cost is low the manager is prone to invest since he knows that he could desinvest at lower prices.

   \subsection{Discretization of the productive assets}

\begin{figure}[H]
\begin{center}
\includegraphics[width = 0.32\textwidth]{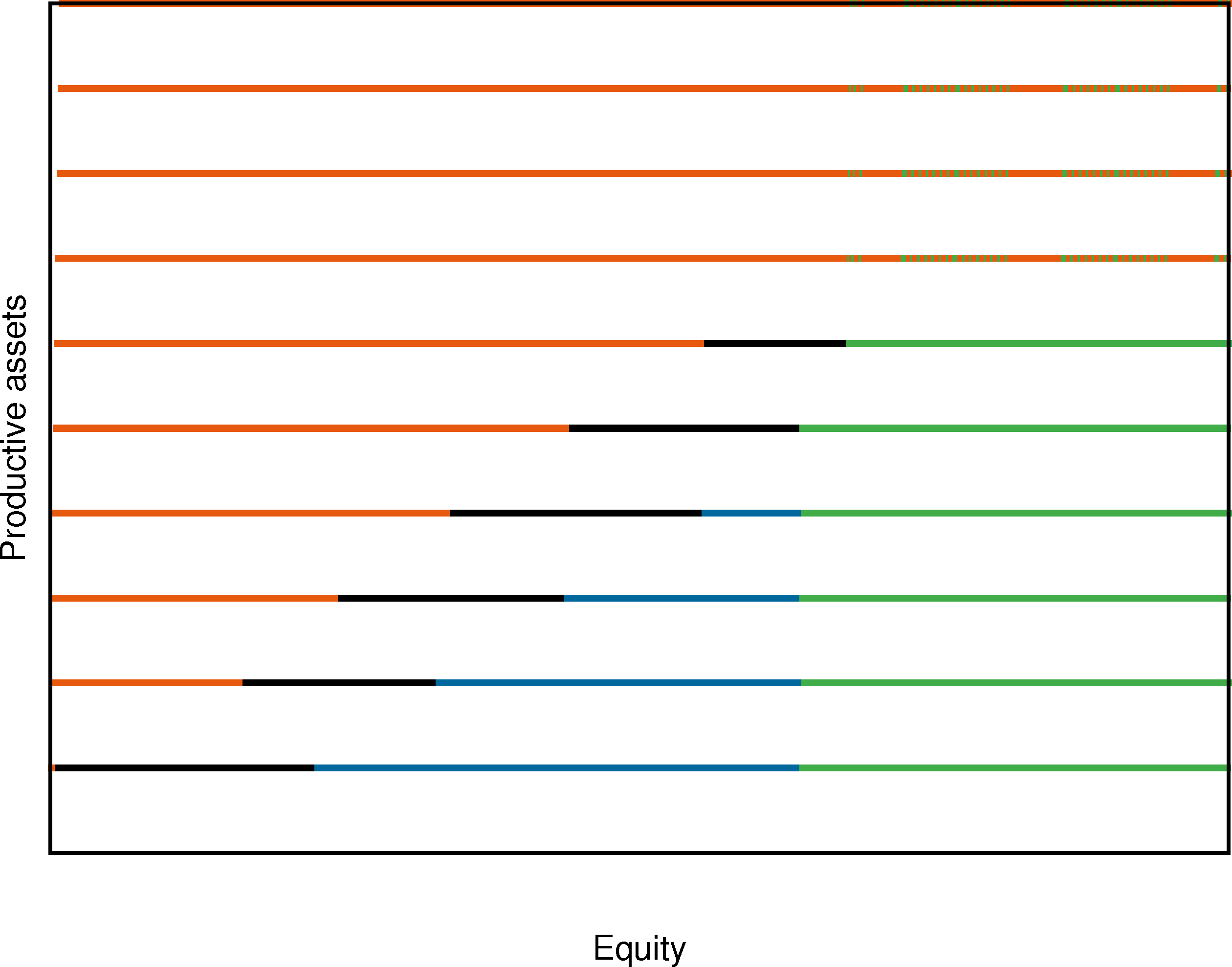}
\includegraphics[width = 0.32\textwidth]{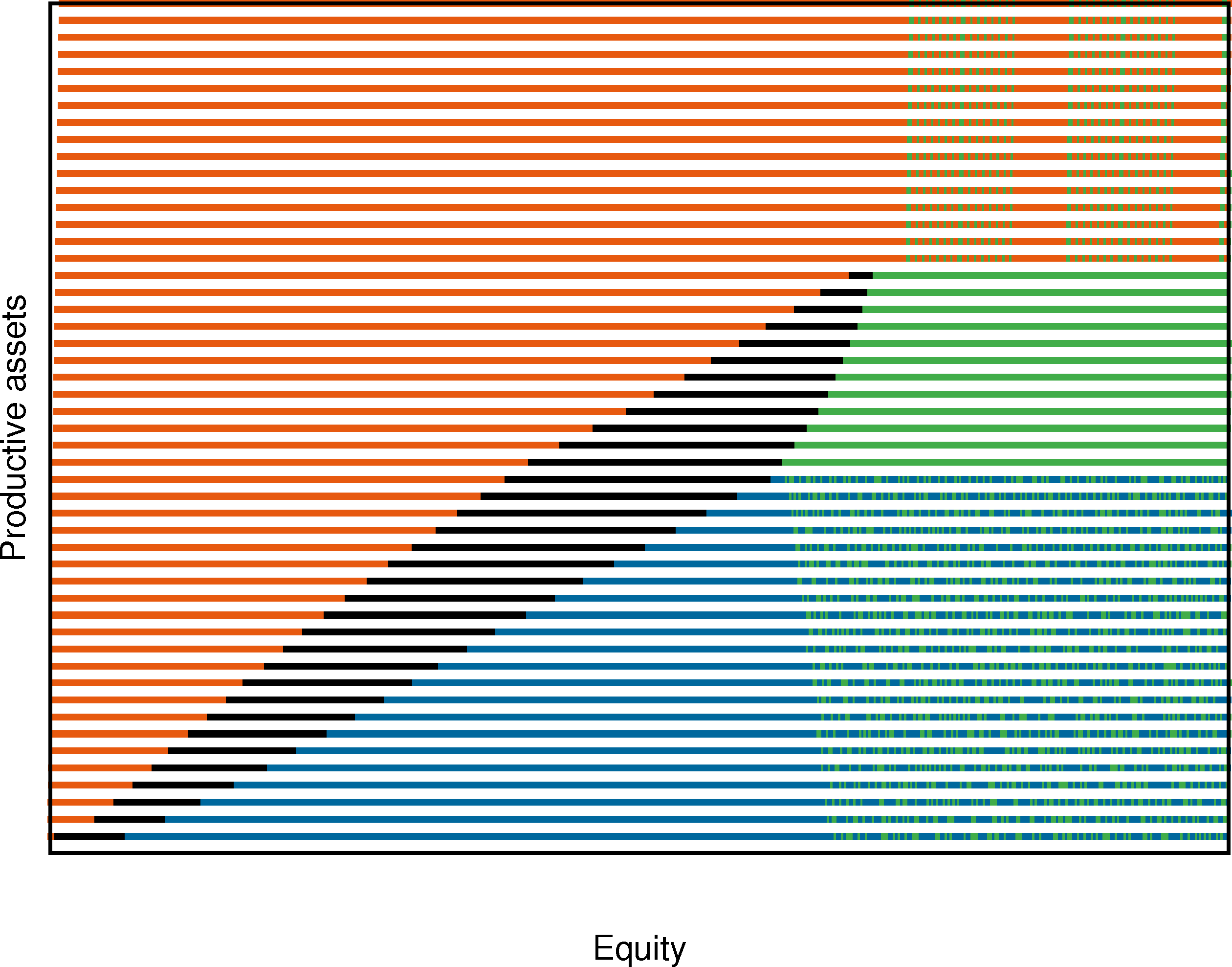}
\includegraphics[width = 0.32\textwidth]{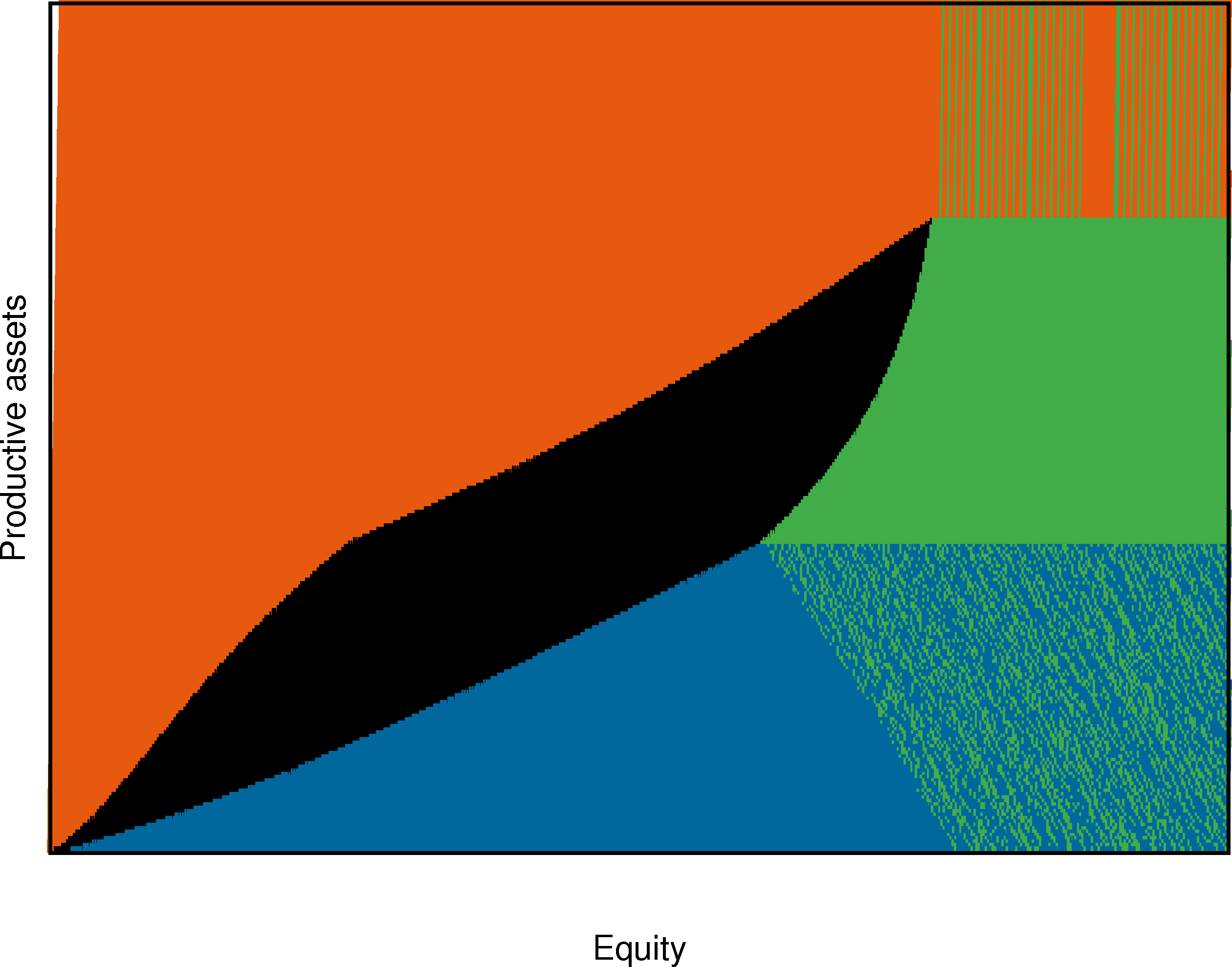}
\caption{Productive assets discretization (k=10, 50 and 250)}\label{opt_k1}
\end{center}
\end{figure}

Above we choose $N=20$ levels of productive assets possible. Figure \ref{opt_k1} shows what happens when we choose a smaller discretization.

\begin{figure}[H]
\begin{center}
\includegraphics[width = 0.32\textwidth]{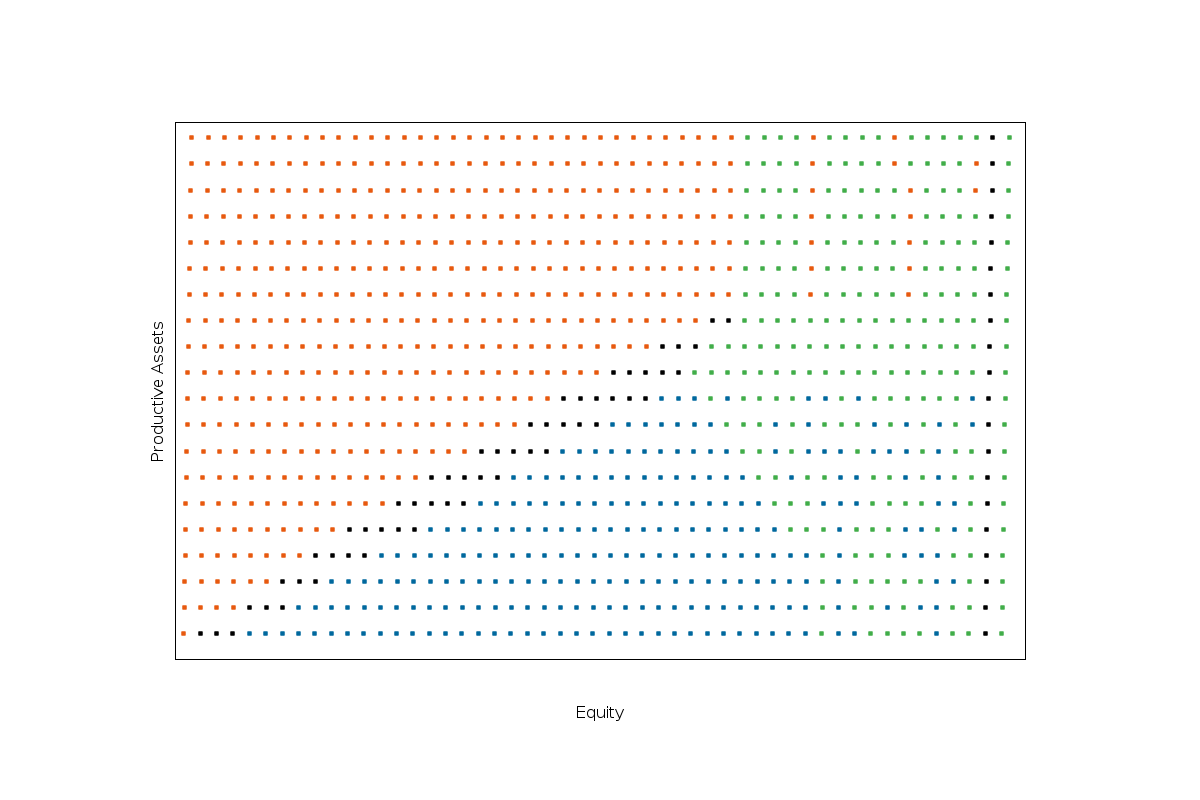}
\includegraphics[width = 0.32\textwidth]{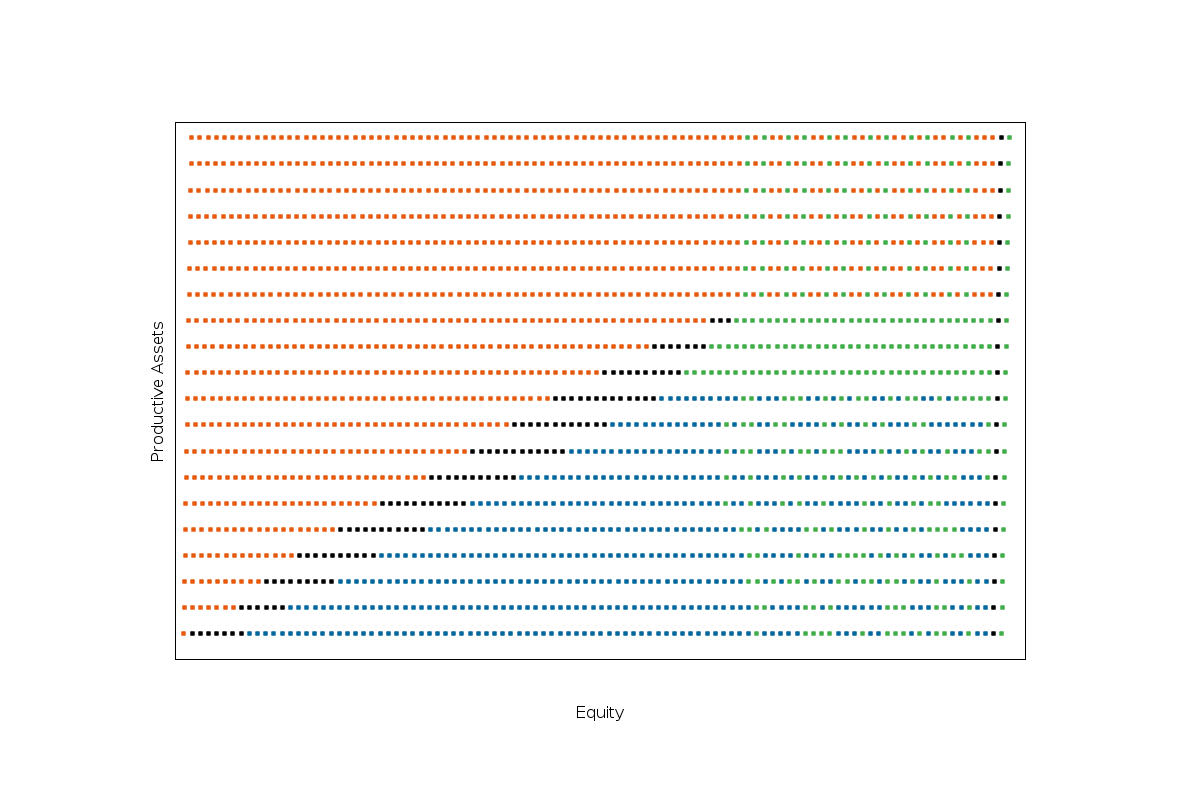}
\includegraphics[width = 0.32\textwidth]{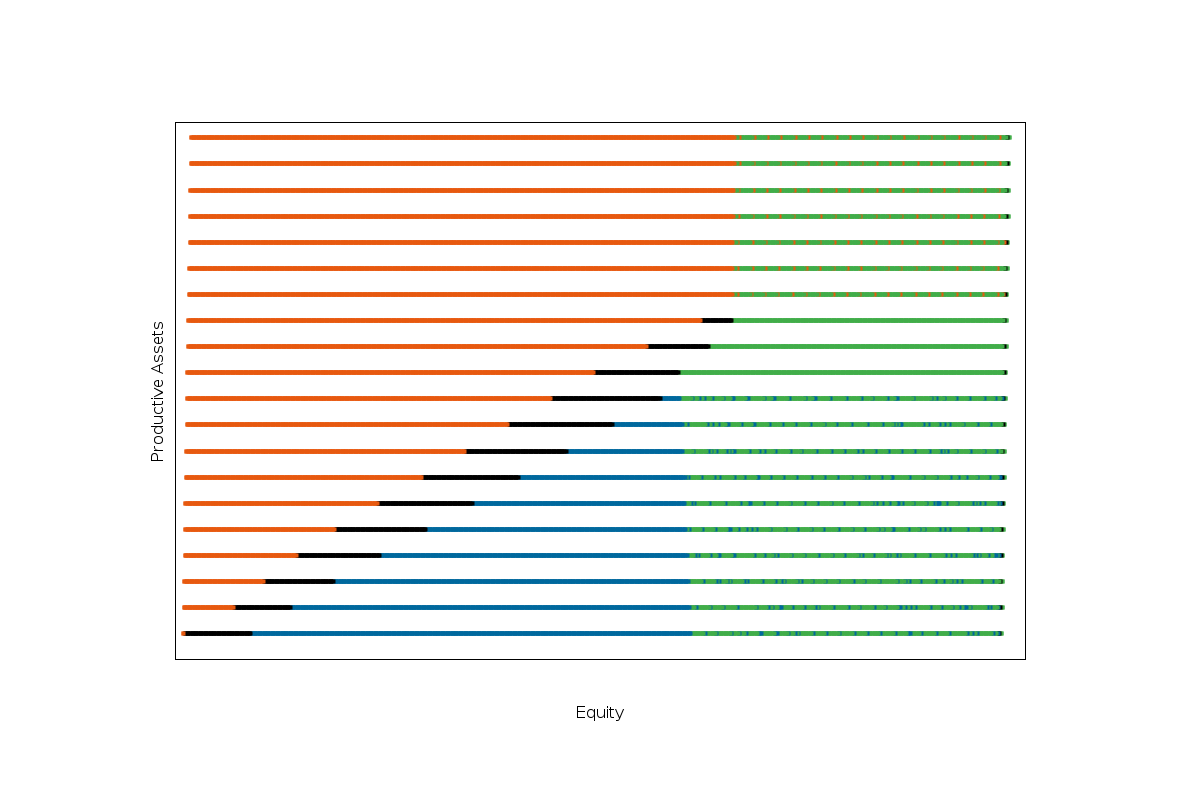}
\caption{Equity discretization (M=50, 100 and 5000)}\label{conv_x}
\end{center}
\end{figure}

Thus, when the productive assets are more liquid, the continuation area (black zone) is wider. In fact, the manager can wait longer before investing or disinvesting because he can do it more often. We also observe the convergence to the true optimal size of the company which is overrun when $N$ is low because of the discretization. The same observation applies to the maximal size of the firm.
To complete the numerical analysis, we present in figure \ref{conv_x} the convergence as $M$ becomes large. We use again $N=20$ for the example.

\section{Conclusion}
The paper describes a financial model of investment for a cash-constrained firm. It derives analytical properties of the value function as well as a description of the shape of the control region. Finally, these theoretical results are used to develop a convergent numerical scheme. The numerical approximation adapts a fixed-point iteration similar to \cite{hfl:12a} for solving the linear system. 

\section{Appendix}

\subsection{Proof of Theorem \ref{conv}}
Using the approximations $q$ and $q+1$, the system (\ref{pol_ite}) can be written as :
\[
\begin{split}
A^{q+1}(U^{q+2}-U^{q+1}) + A^{q+1}U^{q+1} - A^qU^{q+1} &= -B^{q+1}+B^q\\
A^{q+1}(U^{q+2}-U^{q+1}) &=  A^qU^{q+1} + B^q - (A^{q+1}U^{q+1} + B^{q+1}).
\end{split}
\]
We know that $(\rho^{q+1},\psi^{q+1},\theta^{q+1})$ minimize $A(\rho,\psi,\theta)U^{q+1}+B(\rho,\psi,\theta)$ so
\[
A^{q+1}(U^{q+2}-U^{q+1}) \geq 0.
\]
Then using that $A^{q+1}$ is a M-matrix we have
\[
U^{q+2}-U^{q+1} \geq 0.
\]
Therefore the scheme is non-decreasing. Morevover $(A^{q})^{-1}$ and $B^q$ are bounded regardless of $q$ so using that 
\[
U^{q+1} = - (A^q)^{-1}B^q
\]
we know that $U^q$ is also bounded so the scheme is convergent. We note $U^*$ the limit of $(U^q)_{q\in \mathbb{N}}$. We still have to prove that the limit is unique and independent of $U^0$. Suppose there exists $U^*$ and $\bar{U}^*$ two limits. $U^*$ and $\bar{U}^*$ are both solutions of (\ref{eq_lin}) so
\[
\begin{split}
A^*U^* + B^* = 0\\
\bar{A}^*\bar{U}^* + \bar{B}^* = 0
\end{split}
\]
then subtracting the two equations we have,
\[
\bar{A}^*(\bar{U}^* - U^*) = B^* + A^*U^* - (\bar{B}^* + \bar{A}^*U^*).
\]
But $(\bar{\rho}^*,\bar{\psi}^*,\bar{\theta}^*)$ minimize $\bar{A}\bar{U}^* + B$ so
\[
\bar{A}^*(\bar{U}^* - U^*)\leq 0.
\]
Then using that $\bar{A }^*$ is a M-matrix we have
\[
\bar{U}^* - U^*\leq 0.
\]
We prove in the same way that
\[
\bar{U}^* - U^*\geq 0
\]
which achieves the demonstration.

\end{document}